\documentclass[aps,prl,twocolumn,superscriptaddress,longbibliography,nobibnotes]{revtex4-2}

\usepackage{diagbox}
\usepackage{amsmath}
\usepackage{amsfonts}
\usepackage{physics}
\usepackage[space]{grffile}
\usepackage{amssymb}
\usepackage{upgreek}
\usepackage{lineno}
\usepackage{natbib}
\usepackage{braket}
\usepackage{float}
\usepackage{color}
\usepackage{blindtext}
\usepackage{comment}
\usepackage{graphicx}
\usepackage{xr-hyper}
\usepackage{hyperref}
\usepackage{xcolor}
\hypersetup{colorlinks,breaklinks,
            urlcolor=[rgb]{0,0,0.64},
            linkcolor=[rgb]{0,0,0.64},
            citecolor=[rgb]{0,0,0.64},
            filecolor=[rgb]{0,0,0.64}}
\usepackage[T1]{fontenc}

\newcommand{\beginsupplement} {
    \setcounter{table}{0}
    \renewcommand{\thetable}{S\arabic{table}}
    \setcounter{figure}{0}
    \renewcommand{\thefigure}{S\arabic{figure}}
    \setcounter{equation}{0}
    \renewcommand{\theequation}{S\arabic{equation}}
    \setcounter{section}{0}
    \setcounter{secnumdepth}{2}
}

\makeatletter
\def\@ssect@ltx#1#2#3#4#5#6[#7]#8{%
  \def\H@svsec{\phantomsection}%
  \@tempskipa #5\relax
  \@ifdim{\@tempskipa>\z@}{%
    \begingroup
      \interlinepenalty \@M
      #6{%
       \@ifundefined{@hangfroms@#1}{\@hang@froms}{\csname @hangfroms@#1\endcsname}%
       {\hskip#3\relax\H@svsec}{#8}%
      }%
      \@@par
    \endgroup
    \@ifundefined{#1smark}{\@gobble}{\csname #1smark\endcsname}{#7}%
  }{%
    \def\@svsechd{%
      #6{%
       \@ifundefined{@runin@tos@#1}{\@runin@tos}{\csname @runin@tos@#1\endcsname}%
       {\hskip#3\relax\H@svsec}{#8}%
      }%
      \@ifundefined{#1smark}{\@gobble}{\csname #1smark\endcsname}{#7}%
      \addcontentsline{toc}{#1}{\protect\numberline{}#8}%
    }%
  }%
  \@xsect{#5}%
}%
\makeatother

\newcommand{\mytitle}{Dynamical decoding of the competition between charge density waves in a kagome superconductor}

\begin{document}
%TC:ignore
\title{Dynamical decoding of the competition between charge density waves in a kagome superconductor}

\author{Honglie Ning}
\thanks{These authors contributed equally to this work}
\affiliation{Department of Physics, Massachusetts Institute of Technology, Cambridge, MA 02139}

\author{Kyoung Hun Oh}
\thanks{These authors contributed equally to this work}
\affiliation{Department of Physics, Massachusetts Institute of Technology, Cambridge, MA 02139}

\author{Yifan Su}
\thanks{These authors contributed equally to this work}
\affiliation{Department of Physics, Massachusetts Institute of Technology, Cambridge, MA 02139}

\author{Alexander von Hoegen}
\affiliation{Department of Physics, Massachusetts Institute of Technology, Cambridge, MA 02139}

\author{Zach Porter}
\affiliation{Linac Coherent Light Source, SLAC National Accelerator Laboratory, Menlo Park, CA, 94025}
\affiliation{Stanford Institute for Materials and Energy Sciences, Stanford University, Stanford, CA 94305}
\affiliation{Materials Department, University of California, Santa Barbara, CA 93106}

\author{Andrea Capa Salinas}
\affiliation{Materials Department, University of California, Santa Barbara, CA 93106}

\author{Quynh L Nguyen}
\affiliation{Linac Coherent Light Source, SLAC National Accelerator Laboratory, Menlo Park, CA, 94025}

\author{Matthieu Chollet}
\affiliation{Linac Coherent Light Source, SLAC National Accelerator Laboratory, Menlo Park, CA, 94025}

\author{Takahiro Sato}
\affiliation{Linac Coherent Light Source, SLAC National Accelerator Laboratory, Menlo Park, CA, 94025}

\author{Vincent Esposito}
\affiliation{Linac Coherent Light Source, SLAC National Accelerator Laboratory, Menlo Park, CA, 94025}

\author{Matthias C Hoffmann}
\affiliation{Linac Coherent Light Source, SLAC National Accelerator Laboratory, Menlo Park, CA, 94025}

\author{Adam White}
\affiliation{Linac Coherent Light Source, SLAC National Accelerator Laboratory, Menlo Park, CA, 94025}

\author{Cynthia Melendrez}
\affiliation{Linac Coherent Light Source, SLAC National Accelerator Laboratory, Menlo Park, CA, 94025}

\author{Diling Zhu}
\affiliation{Linac Coherent Light Source, SLAC National Accelerator Laboratory, Menlo Park, CA, 94025}

\author{Stephen D Wilson}
\affiliation{Materials Department, University of California, Santa Barbara, CA 93106}

\author{Nuh Gedik}
\email[email: ]{gedik@mit.edu}
\affiliation{Department of Physics, Massachusetts Institute of Technology, Cambridge, MA 02139}

\begin{abstract}
The kagome superconductor CsV$_3$Sb$_5$ hosts a variety of charge density wave (CDW) phases, which play a fundamental role in the formation of other exotic electronic instabilities. However, identifying the precise structure of these CDW phases and their intricate relationships remain the subject of intense debate, due to the lack of static probes that can distinguish the CDW phases with identical spatial periodicity. Here, we unveil the competition between two coexisting $2\cross2\cross2$ CDWs in CsV$_3$Sb$_5$ harnessing time-resolved X-ray diffraction. By analyzing the light-induced changes in the intensity of CDW superlattice peaks, we demonstrate the presence of both phases, each displaying a significantly different amount of melting upon excitation. The anomalous light-induced sharpening of peak width further shows that the phase that is more resistant to photo-excitation exhibits an increase in domain size at the expense of the other, thereby showcasing a hallmark of phase competition. Our results not only shed light on the interplay between the multiple CDW phases in CsV$_3$Sb$_5$, but also establish a non-equilibrium framework for comprehending complex phase relationships that are challenging to disentangle using static techniques.
\end{abstract}
%TC:endignore
\maketitle

\begin{figure*}[t]
\includegraphics[width=6.75in]{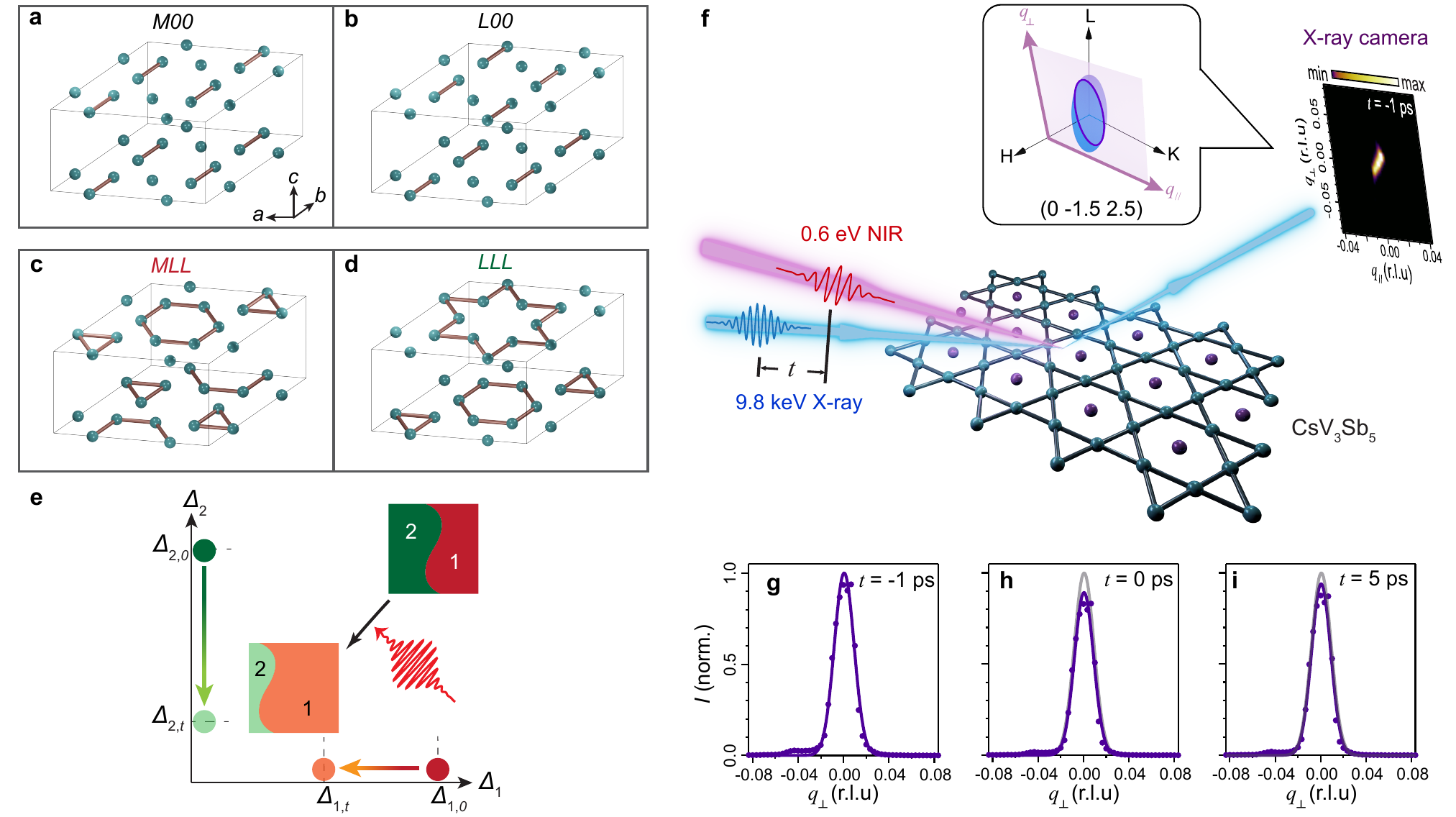}
\label{Fig1}
\caption{\textbf{Phase competition and CDW superlattice dynamics in CsV$_3$Sb$_5$.} \textbf{a,b} Lattice distortions corresponding to the instabilities of phonons at $M$ (0.5 0 0) and $L$ (0.5 0 0.5) points in the momentum space. $M00$ ($L00$) corresponds to the case where the in-plane unit-cell doubling is along the $a$-axis and no doubling is present along the other two symmetry-equivalent directions, namely the $b$- and $(a+b)$-axes. \textbf{c,d} Two most probable $2\cross2\cross2$ CDW structures of CsV$_3$Sb$_5$. Note that in panels \textbf{a-d}, only the contracted bonds are colored in brown and only the V atoms are represented by the green balls for clarity. \textbf{e,} Schematic evolution of the two competing phases, $\Delta_1$ and $\Delta_2$, in the phase space. The red (green) arrow characterizes the change in the amplitude of $\Delta_{1 (2)}$, which is suppressed less (more) by light. Two insets show that phase competition leads to a suppression in order parameter amplitude and a change in real space domain size upon light excitation. \textbf{f,} Schematic of the near-infrared (NIR) pump and hard X-ray probe experimental setup (Methods). Time delay $t$ between the pump and probe can be varied. The sample is rotated nearly around its surface normal to fulfill the Bragg condition for a given diffraction peak. V atoms are represented by green balls while Sb atoms are represented by purple balls. Cs atoms are neglected in the crystal structure for clarity. Raw image of the (0 -1.5 2.5) superlattice peak before pumping is shown. The inset depicts the two-dimensional X-ray camera plane in the three-dimensional momentum space when measuring the peak at (0 -1.5 2.5). The two directions, $q_{\perp}$ and $q_{//}$, are nearly parallel to L and K directions, respectively. This holds true for the other peaks under investigation. \textbf{g-i,} Intensity linecuts $I$ integrated over the $q_{//}$ direction normalized by its equilibrium peak value at representative time delays. Solid lines are fits to Gaussians. Gray curves in panels \textbf{h} and \textbf{i} are the fitting curves reproduced from panel \textbf{g} for better comparison.}
\end{figure*}

Kagome metals $A$V$_3$Sb$_5$ ($A=$ K, Rb, Cs) possess a plethora of proximal phases stemming from the interplay between the non-trivial band topology, lattice geometric frustration, and electronic correlations \cite{WilsonPRM2019,WilsonPRL2020,CominNPHY2022,Jiang2021}. Ensuing or accompanying the emergence of charge density waves (CDWs), a cascade of symmetry breaking phases including orbital flux state \cite{HJPSciBull2021,NeupertPRL2021,FernandesPRB2022,GuguchiaNature2022}, electronic nematicity \cite{CXHNature2022}, and superconductivity accompanied by a pair density wave \cite{BorisenkoPRL2022,WilsonPRM2022,GaulinPRL2022,SatoPRL2022,CominNMAT2023,CXHNature2022} arise. Therefore, deciphering the nature of the CDW phases in this series of compounds is of paramount significance. More intriguingly, unlike the other analogs in this family, CsV$_3$Sb$_5$ hosts a variety of interrelated CDW phases, which exhibit different out-of-plane stacking but identical in-plane $2\cross2$ lattice reconstructions composed of either star-of-David (soD) or inverse-soD patterns \cite{WilsonPRL2020,WilsonPRX2021,GeckPRB2022,PYYPRR2023}. These three-dimensional CDW reconstructions comprise two types of frozen phonons at the $M$ and $L$ points in the momentum space. These $M$ ($L$) modes comprise alternating distortions of V-V bonds that are in-phase (out-of-phase) across neighbouring kagome planes (Fig. 1a,b) \cite{FernandesPRB2021}. Various combinations of the $M$ and $L$ order parameters along the three symmetry-equivalent directions give rise to the rich CDW ecology. Notably, the $2\cross2\cross2$ CDW reconstructions prevail in CsV$_3$Sb$_5$, which compete with the subdominant $2\cross2\cross4$ and $2\cross2\cross1$ stackings \cite{WilsonPRX2021,GeckPRB2022,PYYPRR2023,WilsonPRM2023} and evolve into a quasi-one-dimensional CDW upon doping or under pressure \cite{CXHNature2022_2,WilsonArxiv2022}. 

Despite the intensive investigations, the real-space structure of the dominant $2\cross2\cross2$ CDW ground state has not been conclusively determined. A number of CDW reconstructions with different Landau free energy functionals of the $M$ and $L$ order parameters have been predicted theoretically \cite{FernandesPRB2021}. Different experimental probes have reported contradictory evidence of two possible superlattice configurations, namely either the "$LLL$" structure with alternative aligned soD and inverse soD stacking \cite{SMPRB2022,TjernbergPRR2022,CominNMAT2023,CanosaNCOMM2023}, or the "$MLL$" structure with staggered inverse soD stacking (Fig. 1c,d) \cite{PYYPRR2023,GeckPRB2022,WilsonPRM2023,WLNPHY2022,WNLPRB2022,CXHNature2022,ZeljkovicNature2021}. Although the conflicting experimental observations can be reconciled if the two CDW phases coexist, deterministic evidence demonstrating the presence of both phases has not been uncovered. Moreover, despite the potential coexistence of the two phases, it remains elusive whether they compete or cooperate with each other, hindering the understanding of the other electronic orders.

In general, it is challenging to distinguish the two phases with identical reciprocal lattice vectors using static diffraction-based techniques. However, the relationship between different phases can be uncovered in an out-of-equilibrium pathway. According to a recent theoretical proposal, two competing phases that coexist inhomogeneously in real space will exhibit distinct dynamics upon light stimuli \cite{MillisPRX2020,MillisPRB2020}. Although the amplitude of both phases, characterized by $\Delta_1$ and $\Delta_2$, will be quenched by light irradiation, the domain size of the less suppressed state will expand while the dimension of the other phase will shrink (Fig. 1e). The distinct behaviors of the two phases to light irradiation thus provide direct evidence of phase competition. This non-equilibrium protocol can serve as a smoking gun to distinguish the two charge-ordered phases in CsV$_3$Sb$_5$, because the disparity in their structures could influence the strength of their responses to photo-excitation.

The different responses of the two CDWs can then be uniquely substantiated by time-resolved diffraction: a decrease in CDW satellite peak intensity reflects the suppression of the order parameter amplitude and phase coherence, while the alteration in peak width reveals the change in real-space domain size \cite{Zong2019}. In this work, by interrogating the dynamics of CDW superlattice peak intensity employing time-resolved X-ray diffraction (tr-XRD), we first demonstrate the coexistence of $MLL$ and $LLL$ phases. Furthermore, the CDW peaks contributed by the less suppressed $MLL$ phase exhibit a width decrease owing to the domain expansion, while the peaks contributed by the more suppressed $LLL$ phase display a width increase. These observations unequivocally signify the competition between the two coexisting CDW phases.

\begin{figure*}[t]
\includegraphics[width=6.75in]{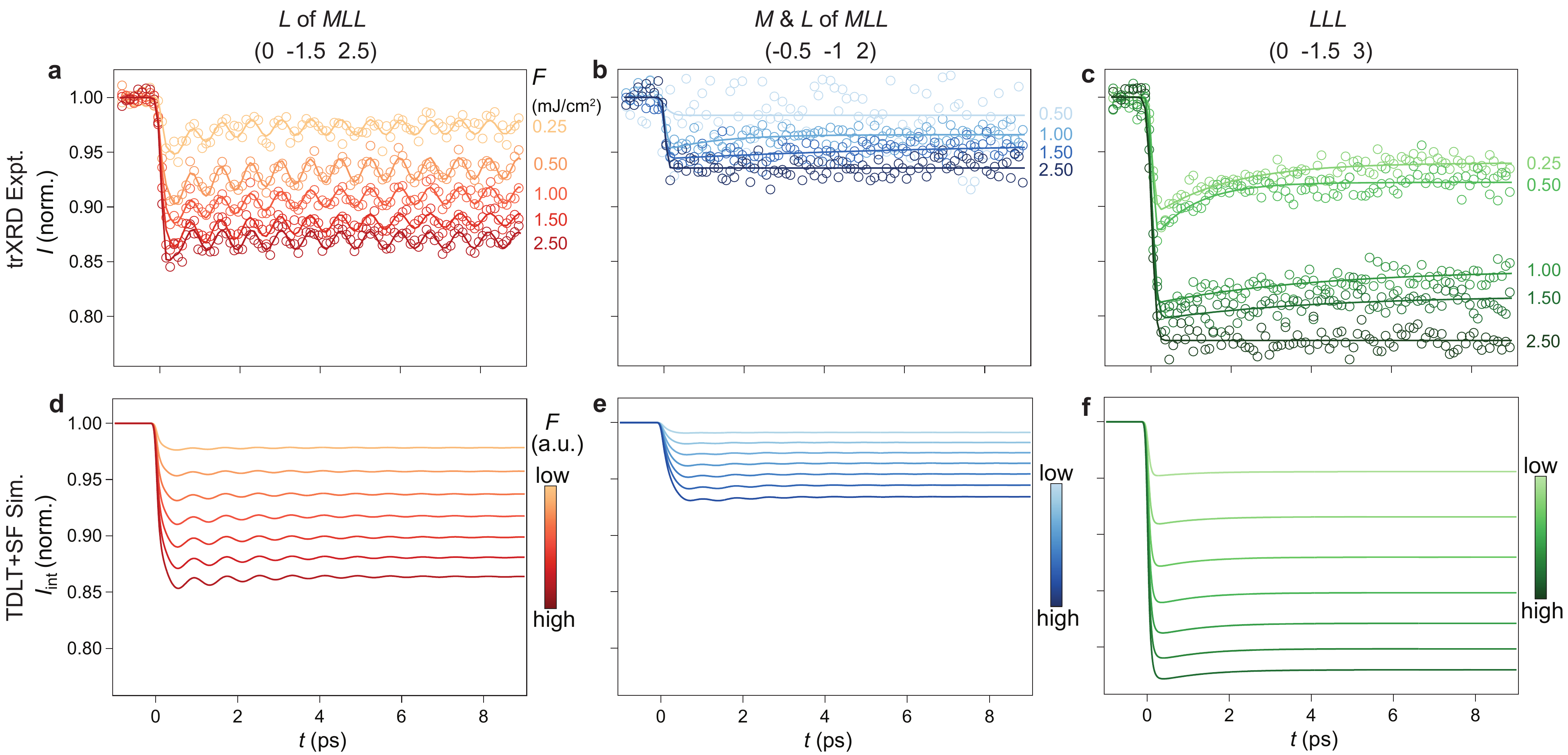}
\label{Fig2}
\caption{\textbf{Dynamics of the CDW peak intensity revealing the coexistence of the CDWs.} \textbf{a-c,} Temporal evolution of the intensity $I$ obtained by tr-XRD measurements at $T$ = 30 K normalized to the static value for select pump fluences $F$ of CDW peaks at (0 -1.5 2.5), (-0.5 -1 2), and (0 -1.5 3), respectively. Solid lines in \textbf{a} are fits to a single damped oscillation atop a single exponential decay, while solid lines in \textbf{b} and \textbf{c} are fits to a single exponential decay (Methods). \textbf{d-f,} Temporal evolution of the normalized integrated intensity $I_{int}$ obtained by simulations based on a combination of time-dependent Landau theory (TDLT) and structural factor (SF) calculation (Methods). The disparity in penetration depths of the X-ray when measuring different peaks has been considered (Supplementary Information Section 13). Fluence ranges are identical for all three peaks. The order parameter governing each peak is shown at the top.}
\end{figure*}

We experimentally investigated three CDW superlattice peaks located at (HKL)\:=\:(0 -1.5 2.5), (-0.5 -1 2), and (0 -1.5 3), where (HKL) denotes the Miller indices. Based on the structure factor calculation results (Methods and Supplementary Information Section 1), we have determined that the first and second peaks are predominantly contributed by the $MLL$ phase, while the third peak can be attributed to both the $MLL$ and $LLL$ phases. Therefore, the ability to observe the first two peaks already demonstrates the existence of $MLL$, consistent with previous static X-ray diffraction results \cite{WilsonPRX2021}. \textcolor{black}{The different peak width of (0 -1.5 3) from (0 -1.5 2.5) hints at its potentially different origin (Supplementary Information Section 9), while more deterministic insights into the potential coexistence of both phases can be provided by} monitoring the dynamics of all three peaks. Moreover, for the peaks generated by the $MLL$ phase, L\:=\:half integer peaks are exclusively caused by $L$ distortion, whereas L\:=\:integer CDW peaks are the result of both $M$ and $L$ distortions. Thus, analyzing both types of peaks will enable us to reveal the dynamics of $M$ and $L$ modes in $MLL$. 

We conducted tr-XRD measurements on single-crystal CsV$_3$Sb$_5$ in the reflection geometry (Fig. 1f and Methods). Diffraction peaks onset at $T_c=91$ K in all three aforementioned positions (Supplementary Information Section 2), signifying the emergence of the long-range CDW \cite{PYYPRR2023,GeckPRB2022,MHPRX2021,MHncomm2022,GaulinPRL2022,CanosaNCOMM2023}. To melt the CDW, we employed an ultrafast laser pulse centered around 0.6 eV as the pump, which excites the electrons around $M$ and $L$ points where CDW instabilities arise \cite{WHHPRB2021}. Moreover, our detection plane is nearly parallel to the (0KL) plane and thus sensitive to changes in both in-plane ($q_{//}$) and out-of-plane ($q_{\perp}$) directions (Fig. 1f inset and Methods). 

We first examine the temporal evolution of the L\:=\:half integer CDW peak at (0 -1.5 2.5) acquired at a temperature of 30 K with a pump fluence of $F=2.5$ mJ/cm$^2$ by varying the time delay $t$ between the pump and the probe. To better visualize the CDW peak profile, we integrate the image along the $q_{//}$ direction and subtract the background (Fig. 1g). At the instant of pump impingement $t=0$, a prompt decrease in the peak intensity can be observed (Fig. 1h). This drop in peak intensity signifies a reduction of $MLL$ amplitude. The rocking curve upon light irradiation shows a uniform decrease in intensity without changing the Bragg condition (Supplementary Information Section 3). At $t=5$ ps, the intensity drop partially recovers (Fig. 1i). A closer examination of the raw image along both $q$-directions reveals that the dynamics is nearly $q$-independent (Supplementary Information Section 4). Hence, we can integrate the image around the superlattice peaks to quantitatively track the temporal evolution of peak intensity $I$ as a function of $t$ at various $F$ (Fig. 2a). Ensuing the rapid decrease in $I$ within 200 fs, which indicates a nonthermal light-induced quenching of the CDW, we observe a partial recovery over several picoseconds to a quasi-equilibrium state that lasts longer than 100 ps. A pronounced oscillation at 1.3 THz can also be resolved atop the recovery process. We tentatively attribute it to a coherent phonon strongly coupled to the CDW, as previously assigned (Supplementary Information Section 5) \cite{HarterPRM2021,WNLPRB2021,GedikArxiv2023}. Note that despite the grazing incidence of the X-ray probe at an angle of 2.94$^\circ$, a large mismatch in the penetration depth of the pump ($\sim$78 nm) \cite{WHHPRB2021} and the probe $\sim$564 nm) renders most of the probed region unexcited. Owing to the approximately 7-fold mismatch, a full melting of CDW at the topmost layer corresponds to a roughly 18\% drop in peak intensity that integrates the entire probed region (Supplementary Information Section 7). Therefore, a maximal drop of 14\% without saturation over $F$ suggests an incomplete melting of $MLL$ of the entire probed region.

In contrast, the L\:=\:integer CDW peak at (-0.5 -1 2), which also derives from the $MLL$ structure, exhibits a significantly smaller light-induced suppression (Fig. 2b). As mentioned before, $L$ distortion should predominantly contribute to the (0 -1.5 2.5) peak, whereas both $M$ and $L$ distortions contribute to the (-0.5 -1 2) peak. The less melting of the latter peak thus suggests that compared to $L$ mode, $M$ mode is more resilient to optical drive, signifying a decoupling of $M$ and $L$ distortions out of equilibrium. This is not entirely surprising as recent X-ray diffraction measurements have uncovered their decoupling under hydrostatic pressure \cite{MHncomm2022}. On the other hand, the L\:=\:integer CDW peak at (0 -1.5 3) shows a substantially larger quenching, which reaches 18\% reduction in $I$ at $F=1.5$ mJ/cm$^2$. Further decrease up to 22\% accompanied by a dynamical slowing down of the recovery time at $F=2.5$ mJ/cm$^2$ suggests the CDW is nearly completely melted in multiple top layers (Fig. 2c and Supplementary Information Section 12) \cite{GedikPRL2019}. If this peak also arises from the $MLL$ phase, a partial melting together with a nearly fluence-independent recovery as the other two peaks should otherwise be expected. This inconsistency can only be reconciled if the (0 -1.5 3) peak \textcolor{black}{mainly} stems from a distinct $LLL$ phase, which is more susceptible than $MLL$ to excitation. Our tr-XRD data thus qualitatively demonstrate the coexistence of $LLL$ with $MLL$. As a reference, Bragg peaks at (0 0 1) and (0 -1 3) show no observable change in the same fluence regime within our signal-to-noise ratio (Supplementary Information Section 6). 

\begin{figure}[t]
\includegraphics[width=3.375in]{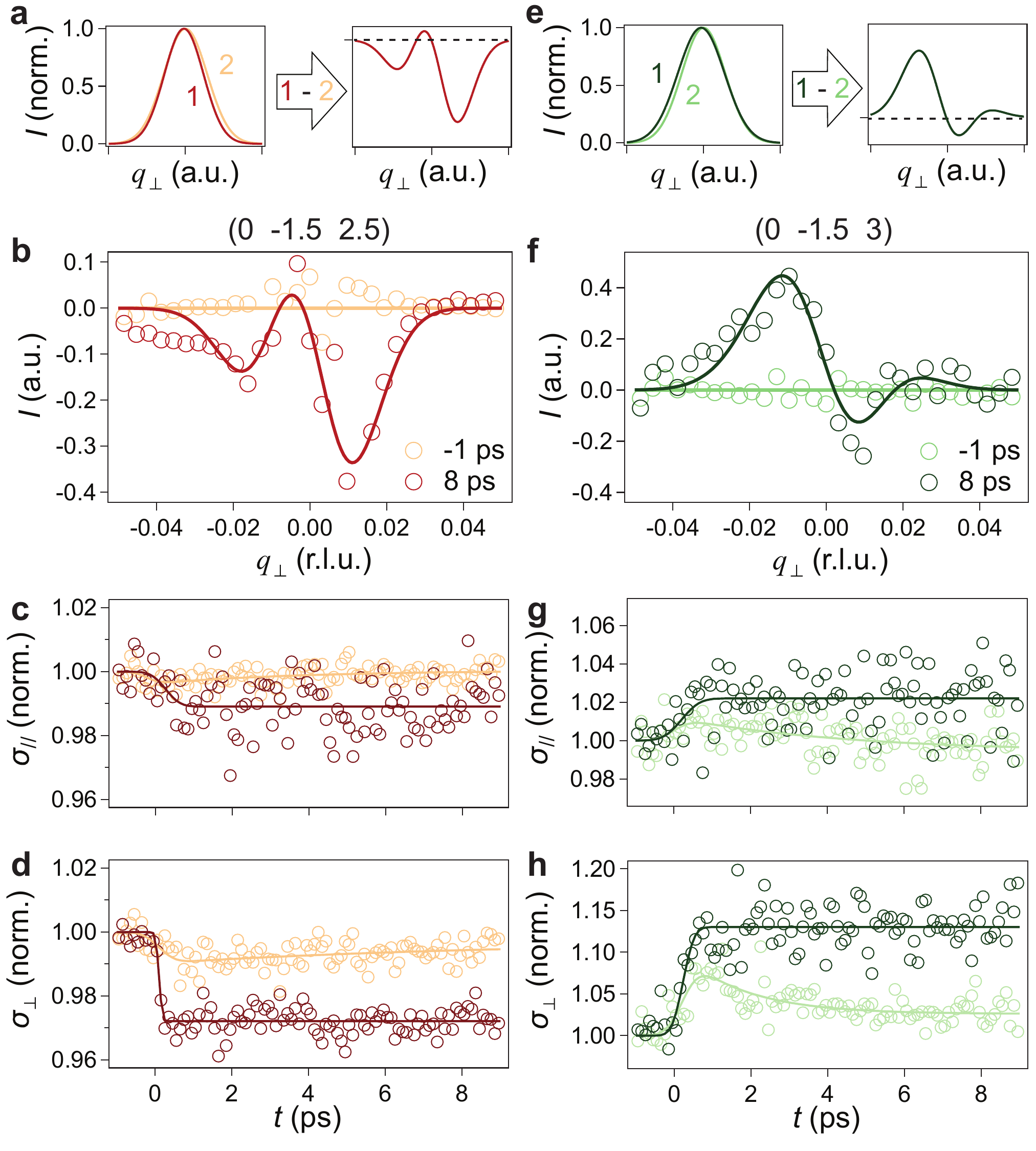}
\label{Fig3}
\caption{\textbf{Dynamics of the CDW peak width revealing the competition between the CDWs.} \textbf{a,} Schematics of the differential peak-intensity-normalized linecuts along $q_{\perp}$, which are obtained by subtracting a broader Gaussian from a narrower Gaussian. \textbf{b,} Differential peak-intensity-normalized linecuts along $q_{\perp}$ acquired at $t=-1$ and $t=8$ ps of the peak at (0 -1.5 2.5). Solid lines are fits to zero constant and subtraction of two Gaussians, respectively. \textbf{c,d,} Temporal evolution of the (0 -1.5 2.5) peak width along $q_{//}$ and $q_{\perp}$ directions pumped at $F=0.25$ (light orange) and 2.5 mJ/cm$^2$ (dark red) normalized by their equilibrium values. Solid lines are fits to a single exponential decay. \textbf{e,} Schematics of the differential peak-intensity-normalized linecuts along $q_{\perp}$, which are obtained by subtracting a narrower Gaussian from a broader Gaussian. \textbf{f-h,} Same plots as panels \textbf{b-d} but for the peak at (0 -1.5 3). Lighter and darker green correspond to pumping at $F=0.25$ and 2.5 mJ/cm$^2$, respectively.}
\end{figure}

To gain a quantitative understanding of the different peak dynamics and examine the coexistence of both CDW phases, we develop a time-dependent Laudau theory (TDLT) \cite{BeaudNMAT2014, trigo2019coherent, JohnsonPRL2014, gorobtsov2021femtosecond, zhou2022dynamical, de2023ultrafast} and leverage it to simulate the temporal evolution of both $M$ and $L$ order parameters, which determines the dynamics of different peaks originating from the $MLL$ and $LLL$ phases. Our dynamical theory is based on a static Landau theory with all the parameters obtained from first-principles calculations (Methods and Supplementary Information Section 7) \cite{YBHPRL2021,FernandesPRB2021,FernandesARXIV2022}. In the time-resolved simulation, we model the light stimulus as a quenching of the quadratic term in the free energy, prone to releasing the CDW distortion and restoring a higher symmetry. However, the quenching of $M$ and $L$ can be different at the same fluence, representing the different susceptibility of $M$ and $L$ order parameters to photo-excitation. We then incorporate the dynamics of $M$ and $L$ modes into the structure factor of different peaks to calculate their intensity evolution. Finally, we include the pump-probe penetration depth mismatch to emulate the experimentally measured data. For the $MLL$ phase, under the assumption that the $M$ mode is approximately 3 times less responsive to photo-excitation than the $L$ mode, we find a quantitative agreement between the simulation and the experiment of the two peaks (Fig. 2d,e). This in turn validates the assumption that $M$ and $L$ exhibit different quenchability and decoupling upon light excitation. The simulated intensity decrease of the (0 -1.5 3) peak, however, shows considerable disagreement with the experimental value (Supplementary Information Section 7). To address this problem, without adding further assumptions, we simulate the dynamics of the $LLL$ phase and find that the fluence required to completely melt it is about 2 times smaller than that of the $MLL$ phase. Its higher susceptibility to light excitation can quantitatively reproduce the larger intensity drop of the peak at (0 -1.5 3) (Fig. 2f). Our simulation hence confirms that the different dynamics of various peaks arise from the coexistence of the two CDW phases that exhibit different susceptibilities to light excitation.

To further investigate whether the $MLL$ and $LLL$ compete with each other, we scrutinize the temporal evolution of the width of various superlattice peaks. To better visualize the subtle light-induced change in peak width, we first normalize the integrated linecuts by their peak values and subtract the linecut before time zero from the one after time zero. By doing so, a decrease in peak width will result in a dip-peak-dip feature, while an increase leads to a peak-dip-peak feature (Fig. 3a,e). For the peak at (0 -1.5 2.5), a dip-peak-dip feature can be clearly resolved after pumping, while the differential linecut is nearly zero before pumping (Fig. 3b). By fitting the peak profile with a Gaussian along both $q_{//}$ and $q_{\perp}$ directions at various $t$, we extract the temporal evolution of the peak full width at half maximum, $\sigma_{//}$ and $\sigma_{\perp}$. Both show a reduction followed by a slow recovery, mirroring the dynamics of the peak intensity, albeit with a delay in rise time (Fig. 3c,d and Supplementary Information Section 13). The temporal evolution of $\sigma_{//}$ and $\sigma_{\perp}$ of the peak at (-0.5 -1 2) exhibit a quantitatively similar behavior, despite a considerably smaller decrease in peak intensity (Supplementary Information Section 8). This similarity supports that the two peaks arise from the same $MLL$ phase whose domain size is expanded by light. This behavior is nonthermal as opposed to the peak broadening around $T_c$ (Supplementary Information Section 9), and different from the conventional observation of light-induced topological defects which cause an increase in peak width \cite{TrigoPRB2021,Zong2019}. On the contrary, the differential linecut, as well as the temporal evolution of $\sigma_{//}$ and $\sigma_{\perp}$ of the peak at (0 -1.5 3), shows an increase in peak width upon pumping (Fig. 3f-h). The opposite behavior of this peak suggests its different origin compared to the other two peaks and reveals the light-induced contraction of the $LLL$ domain. This dichotomy highlights the competition between the $MLL$ and $LLL$ phases, because in the absence of competition, optical quenching will decrease the coherence length of both phases and always induce peak width increase. Another critical observation is that the relative change in $\sigma_{\perp}$ is larger than that of $\sigma_{//}$ for all the peaks, implying optical excitation is more likely to shuffle the out-of-plane stacking rather than reconstructing the in-plane $2\cross2$ distortions. 

\begin{figure}[t]
\includegraphics[width=3.375in]{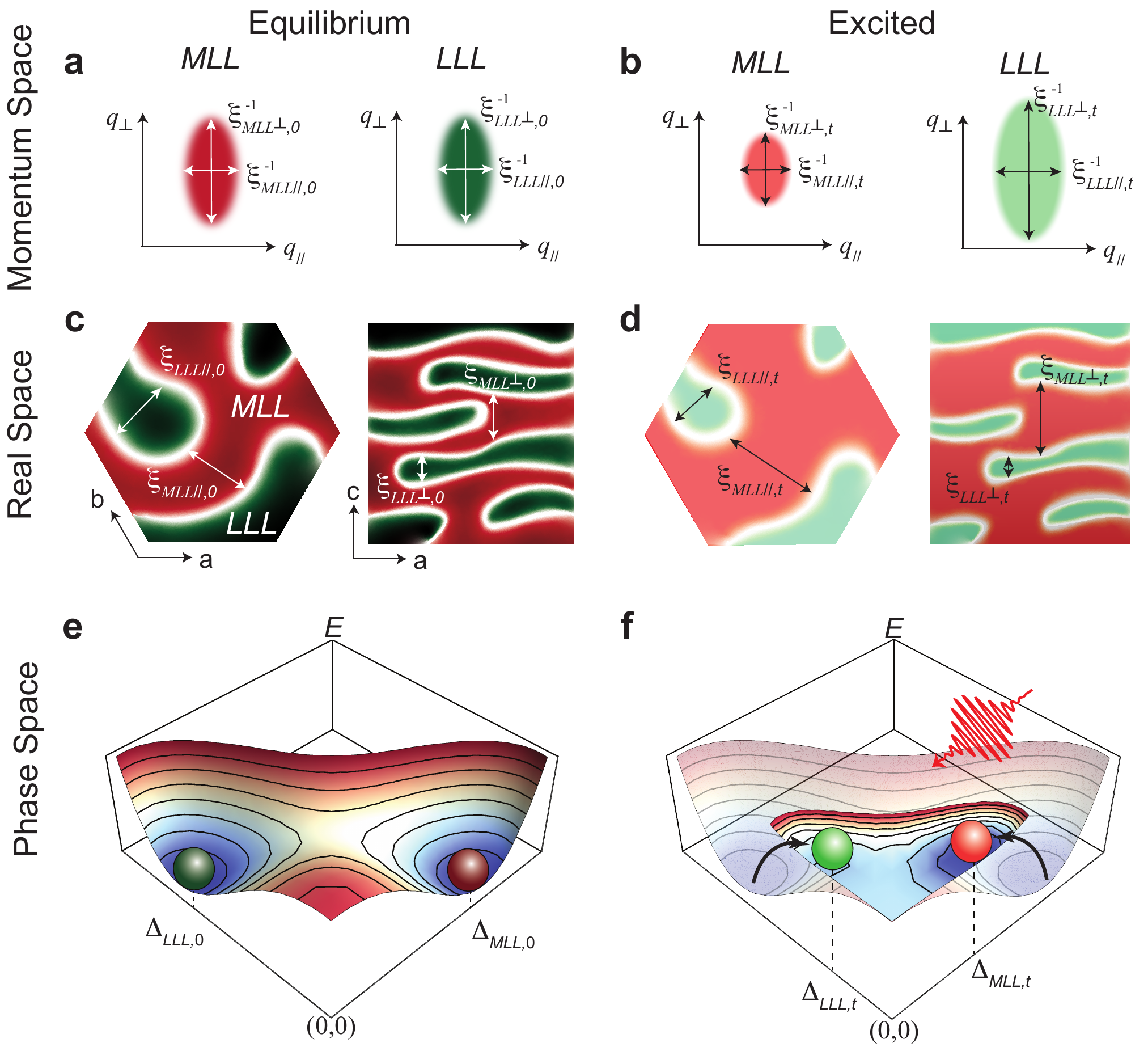}
\label{Fig4}
\caption{\textbf{Physical picture of the CDW dynamics in the presence of phase competition.} \textbf{a,b,} Schematics of equilibrium and excited diffraction patterns of peaks arising from the $MLL$ and $LLL$ structures, respectively. The shallower color in the excited cases characterizes the lower peak intensity, while the change in peak dimensions characterizes the light modulation of peak width. Peak widths are inversely proportional to the correlation lengths as denoted. \textbf{c,d,} Schematics of equilibrium and excited real-space configuration of the $MLL$ and $LLL$ structures in and out of the kagome plane, respectively. Color code is the same as that used in panels \textbf{a} and \textbf{b}. Equilibrium and exited correlation lengths are denoted. \textcolor{black}{The spatial inhomogeneity of light suppression, as reflected by the color gradient, suggests $MLL$ expands the most along both in-plane and out-of-plane directions near the sample surface.} \textbf{e,f,} \textcolor{black}{Schematics of} free energy landscapes of the equilibrium and excited states in the phase space constructed by the $MLL$ and $LLL$ phases (Supplementary Information Section 11). Equilibrium and exited values of the order parameter amplitude corresponding to the two phases are denoted. \textcolor{black}{Note that this figure represents the general case where $LLL$ near the surface is incompletely melted.}}
\end{figure}

Note that the differential linecuts appear asymmetric, because the CDW superlattice peak width change is accompanied by a peak shift, whose amplitude and dynamics are significantly distinct from the weak thermal-expansion-induced Bragg peak shift. Similar phenomena have been observed in photo-induced enhancement of CDW in cuprates \cite{CoslovichScience2022}, stemming from the light-induced change in the phase of the CDW. Using similar methodology, we find a qualitative match between the peak shift value and the peak width change in our compound (Supplementary Information Section 10). Nevertheless, since this displacement is very subtle ($\sim 0.0001$ r.l.u.), we refrain from drawing any quantitative conclusions.

Now we pictorially summarize the physical picture of the optical modulation in Figure 4. Light excitation reduces the peak intensity of both phases and alters the peak width along both in-plane and out-of-plane directions in momentum space (Fig. 4a,b). This change in peak width corresponds to an expansion of the $MLL$ domain and a contraction of the $LLL$ domain along both in-plane and out-of-plane directions in three-dimensional real space (Fig. 4c,d). The opposing dynamics are rooted in a larger suppression of the amplitude of the $LLL$ phase in the phase space (Fig. 4e,f).  

The competition between the two phases can be explained by the fact that both phases tend to open gaps around the Fermi level at similar momentum space locations. Therefore, the loss of density of states resulting from one charge gap inhibits the development of the other. However, the microscopic reasons determining the susceptibility of different CDW phases to excitation cannot be deduced from the phenomenological Landau theory. Based on a recent theory which uncovers that the interlayer coupling plays a crucial role in determining the favored CDW ground state \cite{HYKeeARXIV2023}, we hypothesize that the different melting capabilities of the two CDW phases emerge from their distinct interlayer coupling. Since the out-of-plane distortion of $M$ and $L$ modes has a $\pi$-phase difference, it is intuitive that they will harbor distinct interlayer coupling. This is consistent with our experimental results where $M$ mode is less optically amenable than $L$ mode. Furthermore, this supports the observation that $MLL$ is less quenchable than $LLL$ since it contains $M$ distortion. 

Using a combination of tr-XRD measurements and TDLT simulations, we not only identify the coexistence of two $2\cross2\cross2$ CDW phases in CsV$_3$Sb$_5$, but also illustrate their competing relationship. This discovery raises a wealth of opportunities for scrutinizing the relationship between other CDW reconstructions in this system, such as the $2\cross2\cross4$ \cite{WilsonPRX2021,GeckPRB2022,PYYPRR2023,WilsonPRM2023} and $4\cross1\cross1$ phases \cite{ZeljkovicNature2021}, which both elude our current detection. Understanding the interplay between different CDWs also provides insights for comprehending the unconventional superconductivity in this material, characterized by a pair density wave that inherits the fingerprints of the CDW \cite{GHJNature2021,ZeljkovicARXIV2023,ZXJncomm2022,OkazakiNature2023}. Moreover, this framework for decoding phase competition can be extended to a myriad of exotic electronic orders such as orbital and multipolar orders, which may be detectable by diffraction but remain silent to static external perturbations like magnetic field or pressure. We also envision our work will inspire advances in time-resolved microscopy with atomic resolution, enabling visualization of transient structure changes in three-dimensional real space.

%TC:ignore
\section{Methods}
\subsection{Sample preparation}
Single crystals of CsV$_3$Sb$_5$ were synthesized from Cs (solid, Alfa 99.98\%), V (powder, Sigma 99.9\%, purified with an HCl and Ethanol 9:10 mixture), and Sb (shot, Alfa 99.999\%) using the self-flux method in an argon glovebox with oxygen and moisture levels less than 0.5 ppm. Elemental reagents were loaded into a pre-seasoned tungsten carbide vial to form a precursor flux with a composition of Cs:V:Sb = 20:15:120. Argon-sealed vials were milled for 60 min in a SPEX 8000D mill. The resulting powder was loaded into 2 mL alumina (Coorstek) crucibles and sealed inside steel tubes under argon. Samples were heated to 1000 $^\circ$C at a rate of 200 $^\circ$C/hour, soaked at that temperature for 12 hours, then cooled down to 900 $^\circ$C at a rate of 5 $^\circ$C/hour, and finally slowly cooled to 500 $^\circ$C at a rate of 1 $^\circ$C/hour. Single crystals were extracted mechanically and exfoliated along the [001] direction with scotch tape immediately before the experiment to obtain a fresh and smooth surface. The whole sample was then kept in a vacuum with a pressure lower than $10^{-6}$ torr at low temperature.

\subsection{Time-resolved X-ray diffraction measurements}
The experiments were performed at the X-ray pump-probe (XPP) end station of the Linac Coherent Light Source (LCLS), SLAC National Accelerator Laboratory \cite{Emma2010, Chollet:yi5004}. An optical parametric amplifier seeded by a Ti:Sapphire laser produces a near infrared (NIR) laser pulse centered around 0.6 eV with a duration of 50 fs, which was used as a pump. It excited the sample at an incident angle of 13$^\circ$ with p-polarization and with a focal spot size of $0.27\cross1.2$ mm$^2$. The fluence of the pump is tunable from 0 to 4 mJ/cm$^2$, above which thermal damage may occur. A time-delayed X-ray pulse centered around 9.835 keV with a duration of 40 fs was employed as the probe. It was incident at a grazing angle of 2.95$^\circ$ with respect to the sample surface and with a focal spot size of $0.1\cross0.25$ mm$^2$. The single-shot diffraction patterns were recorded on a two-dimensional Jungfrau1M detector positioned 155 mm away from the sample. The timing jitter between the pump and probe beams was corrected on a single-shot basis using a spectroencoding diagnostic tool to achieve a time resolution of $<80$ fs. 

During the experiment, we first scanned the rocking curve around the Bragg position of each peak at critical time delays: one before time zero  ($t = -2$ ps) and another when the light-induced change is maximal ($t = 0.5$ ps). The peak intensity uniformly decreases without noticeable peak shift, indicating a predominant melting of the CDW order. Therefore, to track the intensity change, we can measure the time traces at the optimized Bragg condition —  at the peak of the rocking curves — without the need to measure rocking curves at every time delay.

After acquiring the CCD images, the first step in our analysis involves subtracting the background intensity, calculated by averaging the intensity of pixels far from the diffraction peaks on the CCD, from each image. This background subtraction is performed at every time delay, ensuring that the relative intensity change $\Delta I/I$, as shown in Fig. 2, arises exclusively from the peak intensity change. Therefore, the difference in intensity suppression among peaks is not artificially influenced by background intensity.

\subsection{Structure factor simulations}
We performed the structure factor (SF) calculation to simulate the peak intensity. Without considering any CDW-induced lattice distortion, SF reads
\begin{equation*}\label{SF}
    \sqrt{I}=S_{\text{HKL}}=\Sigma_j f_j \exp[2\pi i (\text{H}x_j+\text{K}y_j+\text{L}z_j)],
\end{equation*}
where H, K, L are Miller indices for a diffraction peak, $x_j$, $y_j$, $z_j$ are positions of $j$-th atom in a unit cell, and $f_j$ is an atomic form factor depending on (HKL). Here, we only consider in-plane V-V bond distortions within the two kagome planes, because they are the leading lattice instabilities that generate the CDW as predicted by theories and demonstrated by X-ray diffraction measurements \cite{FernandesARXIV2022,WilsonPRM2023}. The modified SF due to the CDW formation can be easily calculated by counting all the V atoms in the expanded unit cell and substituting $x_j\rightarrow x_j+\delta x_j$ and $y_j\rightarrow y_j$+$\delta y_j$, where $\delta x_j$ and $\delta y_j$ correspond to the distortion induced by the V-V bond contraction. See Supplementary Information Section 1 for more details.

\subsection{Time-dependent Landau theory simulations}
The equilibrium Landau free energy functional of the $MLL$ and $LLL$ phases has been derived elsewhere \cite{FernandesPRB2021,FernandesARXIV2022} and all the Landau parameters have been determined by density functional theory simulations. Both potentials can be expanded up to the quartic order of $M$ and $L$ order parameters. To account for the phonon oscillation exclusively occurring in L\:=\:half integer peaks, we include a lattice order parameter $X$ in a harmonic potential which primarily linearly couples to $L$. Therefore, the total free energy is the summation of two parts (see Supplementary Information Section 7 for a detailed derivation):
\begin{equation*}\label{FreeEnergy}
    F=F_{MLL/LLL}+F_X
\end{equation*}
\begin{equation*}\label{FreeEnergyMLL}
\begin{split}
    F_{MLL}=&\frac{\alpha_M(1-T/T_M)}{2}M^2+\frac{u_M}{4}M^4+\\
    &\frac{2\alpha_L(1-T/T_L)}{2}L^2+\frac{4u_L+\lambda_L}{4}L^4+\\
    &\frac{\beta_{ML}}{3}ML^2+\frac{2\lambda_{ML}^{(3)}}{4}M^2L^2,
\end{split}
\end{equation*}
\begin{equation*}\label{FreeEnergyLLL}
\begin{split}
    F_{LLL}=&\frac{3\alpha_M(1-T/T_M)}{2}M^2+\frac{\beta_M}{3}M^3+\frac{9u_M+3\lambda_M}{4}M^4+\\
    &\frac{3\alpha_L(1-T/T_L)}{2}L^2+\frac{9u_L+3\lambda_L}{4}L^4+\\
    &\frac{3\beta_{ML}}{3}ML^2+\frac{3\lambda_{ML}^{(1)}+3\lambda_{ML}^{(2)}+9\lambda_{ML}^{(3)}}{4}M^2L^2,
\end{split}
\end{equation*}
and
\begin{equation*}\label{FreeEnergywithLattice}
    F_X=\frac{1}{2}\omega^2X^2+gXL,
\end{equation*}
where $\alpha_M$ and $\alpha_L$ are the coefficients of the quadratic terms with tunable temperature $T$ and different transition temperatures $T_M$ and $T_L$. Both of them should be negative so that the ordered phase can be realized when $T<T_{M/L}$. $u_M$, $u_L$, $\lambda_M$, $\lambda_L$, $\lambda_{ML}^{(1)}$, $\lambda_{ML}^{(2)}$, and $\lambda_{ML}^{(3)}$ are the coefficients of the quartic terms. $\beta_M$ and $\beta_{ML}$ are the coefficients of the trilinear coupling terms. $\omega$ is the coupled phonon frequency, and $g$ is the electron-phonon coupling constant.

We assume $M$ and $L$ are electronic and thus display overdamped dynamics, while $X$ displays an underdamped dynamics. The dynamical equations of order parameters can be expressed as (See Supplementary Information Section 7 for detailed explanations):
\begin{equation*}\label{Dynamicalequ}
\begin{split}
    \frac{1}{\gamma_{M}}{{\partial}_t}{M}&=-\frac{\partial F}{\partial M},\\
    \frac{1}{\gamma_{L}}{{\partial}_t}{L}&=-\frac{\partial F}{\partial L},\\
    {{\partial}_t}^2{X}=-&2\gamma_X{\partial}_t{X}-\frac{\partial F}{\partial X},
\end{split}
\end{equation*}
where $\gamma_M$, $\gamma_M$, and $\gamma_X$ are phenomenological decay constants to account for damping of electronic and structural order parameters. 

Temporal evolution of $M$ and $L$ are calculated by numerically solving the equations of motion. Their dynamics are then included in the structure factor calculations to simulate the dynamics of different CDW peaks.

\subsection{Fitting formulas for the time traces}
To fit the temporal evolution of peak intensity and peak width, we used the following expression including a single exponential decay, a single damped oscillation, and a background: 
\begin{equation*}\label{FittingFormula}
    \frac{1}{2}[1+\ \mathrm{erf}(\frac{t-t_0}{\sqrt{2}t_r})]\times[Ae^{-\frac{t-t_0}{T}}+Be^{-\frac{t-t_0}{\tau}}\cos(\omega t+\phi)+C],
\end{equation*}
\noindent
where $t$ is the pump-probe time delay, $t_{0}$ is time zero, $t_{r}$ is the rise time, $A$ is the amplitude of the exponential decay, $T$ is the decay time, $B$ is the amplitude of the oscillation, $\tau$ is the oscillation decay time, $\omega$ is the oscillation angular frequency, $\phi$ is the phase, and $C$ is the offset on the measured timescale. The phonon term is only included for the fitting in Fig. 2a.  

\section{Acknowledgements}
The authors thank Edoardo Baldini, Riccardo Comin, Pavel Dolgirev, Emre Erge\c{c}en, Mingu Kang, Anshul Kogar, Tianchuang Luo, Zhiyuan Sun, Zhuquan Zhang, Minhui Zhu, and Alfred Zong for stimulating discussions. The work at MIT was supported by the National Science Foundation under Grant No. DMR-2226519 (data analysis and modeling) and Gordon and Betty Moore Foundation’s EPiQS Initiative grant GBMF9459 (data taking). A.v.H. gratefully acknowledges funding by the Alexandar von Humboldt foundation. S.D.W and A.C.S. gratefully acknowledge support via the UC Santa Barbara NSF Quantum Foundry funded via the Q-AMASE-i program under award DMR-1906325. Z.P is supported by the U.S. Department of Energy, Office of Science, Office of Basic Energy Sciences, under contract DE-AC02-76SF00515, both through the Division of Materials Sciences and Engineering and through the use of the LCLS, SLAC National Accelerator Laboratory. Q.L.N. acknowledges the Bloch Fellowship in Quantum Science and Engineering by the Stanford-SLAC Quantum Fundamentals, Architectures and Machines Initiative. Use of the LCLS, SLAC National Accelerator Laboratory, is supported by the U.S. Department of Energy, Office of Science, Office of Basic Energy Sciences under Contract No. DE-AC02-76SF00515.

\section{Author Contributions}
H.N., K.H.O., Y.S., A.v.H., Z.P., A.C.S., Q.L.N., M.C., and T.S. performed the experiment. A.C.S. and S.D.W. synthesized, characterized, and prepared the samples for the experiment. Q.L.N., T.S., M.C.H., A.J.W., C.M., and D.Z. maintained the beamline and set up the accompanying optics used in the experiment. H.N. performed theoretical calculations. H.N., K.H.O., Y.S., A.v.H., Z.P., and A.C.S. analyzed the data with help from Q.L.N. and V.E. H.N., K.H.O., and Y.S. wrote the manuscript with critical input from D.Z., S.D.W., N.G, and all other authors. The work was supervised by N.G.

\section{Competing Interests}
The authors declare no competing interests.

\section{Data Availability}
The data that support the plots within this paper and other findings of this study are available from the corresponding author upon reasonable request.

\providecommand{\noopsort}[1]{}\providecommand{\singleletter}[1]{#1}%
%

%%%%%%%%% Supplementary Information %%%%%%%%%%%%%%
\clearpage
\newpage

\onecolumngrid
\begin{center}
\textbf{\large Supplemental Material to ``\mytitle''}
\end{center}\hfill\break
\twocolumngrid

\beginsupplement

\makeatletter
\let\toc@pre\relax
\let\toc@post\relax
\makeatother 

\section*{S1. Structure factor calculations}

In order to simulate the peak intensity, we calculate the structure factor (SF), which is proportional to the square root of the peak intensity. Without considering any charge density wave (CDW)-induced lattice distortion, the SF $S$ can be expressed as:
\begin{equation}\label{SF}
    S=\Sigma_j f_j \exp(2\pi i (\text{H}x_j+\text{K}y_j+\text{L}z_j)),
\end{equation}
where H, K, L are Miller indices for a diffraction peak, $x_j$, $y_j$, $z_j$ are the positions of the $j$-th atom in a unit cell, and $f_j$ is an atomic form factor that depends on (HKL). Here, we only consider in-plane V-V bond distortions within the two kagome planes, because they are the leading lattice instabilities that generate the CDW as predicted by theories and demonstrated by X-ray diffraction measurements \cite{FernandesARXIV2022,WilsonPRM2023}. 

As mentioned in the main text, the two candidates for the $2\cross2\cross2$ CDW structure are the $MLL$ (and its symmetry-equivalent nematic domains, i.e. $LML$, $LLM$) and $LLL$ structures, formed by a combination of one $M$ distortion and two $L$ distortions and three $L$ distortions along three directions 120$^\circ$ apart from each other, respectively. The modified SF due to the CDW formation can be easily calculated by counting all the V atoms in the expanded unit cell and substituting $x_j\rightarrow x_j+\delta x_j$ and $y_j\rightarrow y_j$+$\delta y_j$, where $\delta x_j$ and $\delta y_j$ correspond to the distortion induced by the V-V bond contraction. In this simulation, we assume that the relative V-V bond distortion is about 0.5\% compared to the undistorted bond length for all $M$ and $L$ distortions \cite{BlumbergPRB2022}. The simulation results of the CDW superlattice peak intensity in the (0 K L) plane are shown in Fig. S1. Although both $MLL$ and $LLL$ phases can plausibly create all half-integer (H K L) peaks since both structures induce a $2\cross2\cross2$ unit cell expansion, the SF calculation shows that in the $LLL$ structure, the peak intensity at (0 -1.5 2.5) (red circle) accidentally vanishes as the SF contribution from the upper and lower kagome plane cancels each other. On the other hand, the peak intensity at (0 -1.5 3) (green circle) in the $LLL$ structure is similar to that of the $MLL$ structure, so this peak could originate from both the $MLL$ and $LLL$ structures. Lastly, the peak at (-0.5 -1 2) is also predominantly contributed by the $MLL$ structure, albeit not shown here. The above statement is qualitatively true even if we consider the $LLL$ (85\%)+$MMM$ (15\%) structure evidenced by recent ARPES experiments \cite{CominNMAT2023} rather than the pure $LLL$ structure: the peak intensity at (0 -1.5 2.5) and (-0.5 -1 2) are at least two orders of magnitude smaller in $LLL+MMM$ than in $MLL$, while the intensity at (0 -1.5 3) is on the same order for both structures. \textcolor{black}{Note that $LLL+MMM$ is actually a single homogeneous phase without domain structures. It represents a structure very similar to $LLL$, with the only disparity being that the amount of distortion in the inverse star of David and star of David is different. One could also argue that the proportion of the $LLL$ phase may be several orders of magnitude larger than that of the $MLL$. However, this is inconsistent with any previous static X-ray diffraction measurements, which all provide evidence of $MLL$. If $LLL$ were dominant, X-ray should manifest features of its presence. We therefore conclude that (0 -1.5 2.5) and (-0.5 -1 2) indeed are predominantly contributed by $MLL$.} Since the three CDW peaks investigated in this work may originate from different CDW structures, it is possible to track the responses of various CDW domains by monitoring different CDW peaks.

\begin{figure*}[tb!]
\centering
\includegraphics[width=0.8\textwidth]{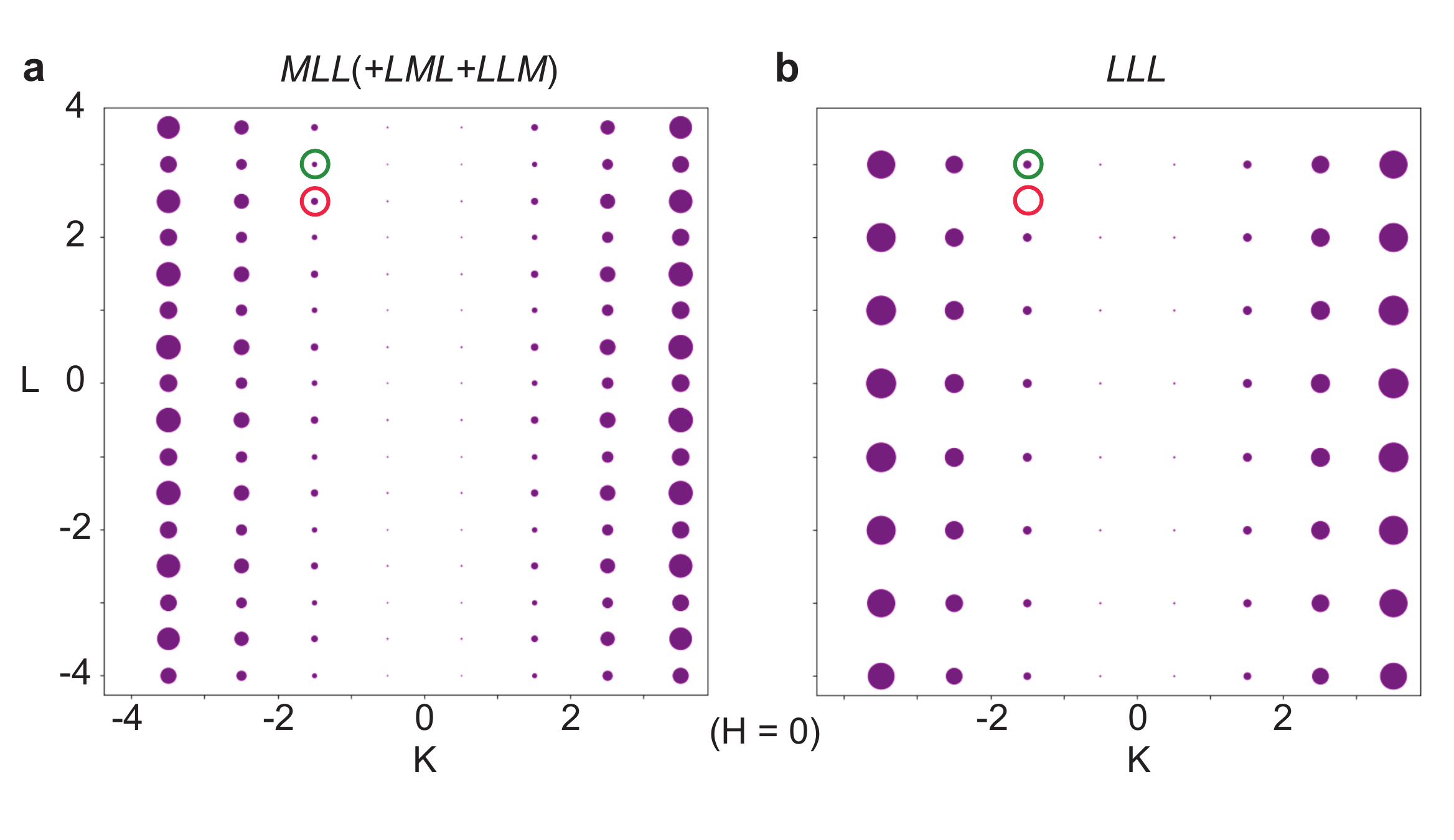}
\label{FigS1}
\caption{\textbf{Structure factor calculation results for CDW satellite peaks.} Calculated (0 K L) CDW superlattice peak intensity map of \textbf{a}, the $MLL$ (and its symmetry-equivalent nematic domains $LML$, $LLM$) and \textbf{b}, the $LLL$ structures. The red and green circles denote the peaks at (0 -1.5 2.5) and (0 -1.5 3.0) studied in this work, respectively. We neglect the Bragg peaks for simplicity.}
\end{figure*}

\section*{S2. Temperature dependence of the CDW peak intensity}

We first scrutinize the static CDW peak intensity as a function of temperature $T$ close to the CDW transition temperature $T_c$ at the three peak positions (0 -1.5 2.5), (-0.5 -1 2), and (0 -1.5 3). We observe that nonzero intensity at the CDW reciprocal lattice vectors only appears below $T_c=91$ K. We note that the peak intensity is only one to two orders of magnitude larger than the background given our experimental geometry and accumulation time, demonstrating the relatively weak CDW peak intensity \cite{PYYPRR2023}. Above $T_c=91$ K, we detect no diffuse scattering at these three peak positions within our signal-to-noise ratio.

We note that although previous X-ray diffraction studies have reported the temperature dependence of selected CDW peaks, their temperature step is usually on the order of 1 K or larger \cite{PYYPRR2023,GeckPRB2022,MHPRX2021,MHncomm2022,GaulinPRL2022,CanosaNCOMM2023}. In contrast, we conduct a temperature scan with a finer step of 0.1 K and within a narrower range around $T_c$, allowing us to better resolve the critical behavior of different peaks. We also note that the Bragg conditions which are fulfilled at 88 K almost hold for the entire temperature range. Our careful measurements show that the L\:=\:half integer peaks exhibit a smoother intensity onset as temperature decreases compared to the two L\:=\:integer peaks (Fig. S2). As discussed in the main text, the former is mainly contributed by the $L$ order parameter, while the latter peaks are contributed by both $L$ and $M$ order parameters. As later shown in Section 7, $L$ exhibits a second-order phase transition while $M$ depicts a first-order transition at $T_c$, when they are not coupled (insets of Fig. S2). These differences provide an explanation for the distinct critical temperature dependence around $T_c$ of different peaks with different indices and lay the foundation of their decoupling upon pumping.

We then unblock the pump and measure the peak intensity at different temperatures at a time delay $t=0.5$ ps with a pump pulse at $F=1.5$ mJ/cm$^2$. The peaks at (0 -1.5 2.5) and (0 -1.5 3) show an obvious intensity drop upon pumping at all temperatures below $T_c$ (Fig. S2a and S2c), with relative amplitudes similar to those measured at 30 K. The change in intensity of the peak at (-0.5 -1 2) is less apparent (Fig. S2b), also in agreement with its much smaller change at 30 K compared to the other two peaks. 

It is worth noting that $T_c$ does not change within our experimental resolution even when the pump laser is turned on, indicating negligible pump-induced heating at least around $T_c$. Since the heat capacity is just three times smaller at 30 K compared to that at 90 K \cite{WilsonPRL2020}, the heating at 30 K should be three times larger than that at 90 K and thus also marginal. 

\begin{figure*}[!htbp]
\includegraphics[width=1\textwidth]{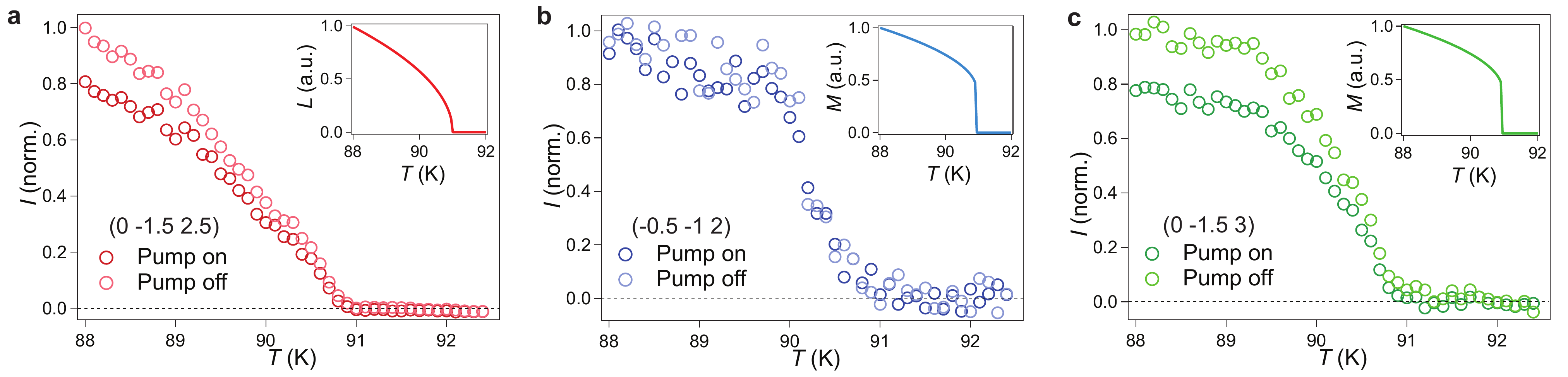}
\label{FigS2}
\caption{\textbf{Temperature-dependence of the integrated intensity of CDW peaks near $T_c$.} The integrated intensity as a function of temperature for \textbf{a}, the peak at (0 -1.5 2.5), \textbf{b}, the peak at (-0.5 -1 2), and \textbf{c}, the peak at (0 -1.5 3) both in the presence and absence of a pump at $F=1.5$ mJ/cm$^2$ acquired at $t=0.5$ ps. Landau theory simulations of the temperature dependence of $L$ (second-order) and $M$ (first-order) are depicted in the inset.}
\end{figure*}

\section*{S3. Rocking curve scans with and without pump}

We scanned the rocking curve around the Bragg position of each peak by rotating the $\mu$ angle of the six-circle diffractometer \cite{You1999} with a 0.02$^\circ$ step to find the optimal Bragg condition that maximizes the peak intensity. As shown in Fig. S3, upon light impingement, the peak intensity uniformly decreases without any noticeable peak shift, indicating a predominant melting of the CDW order. This holds true for all the CDW satellite peaks investigated throughout the experiment. The uniform intensity drop across the $\mu$ range indicates that the significant difference in the intensity drop for the three CDW peaks is not an artifact caused by light-induced deviations from the exact Bragg condition.

\begin{figure}[!htbp]
\centering
\includegraphics[width=0.9\columnwidth]{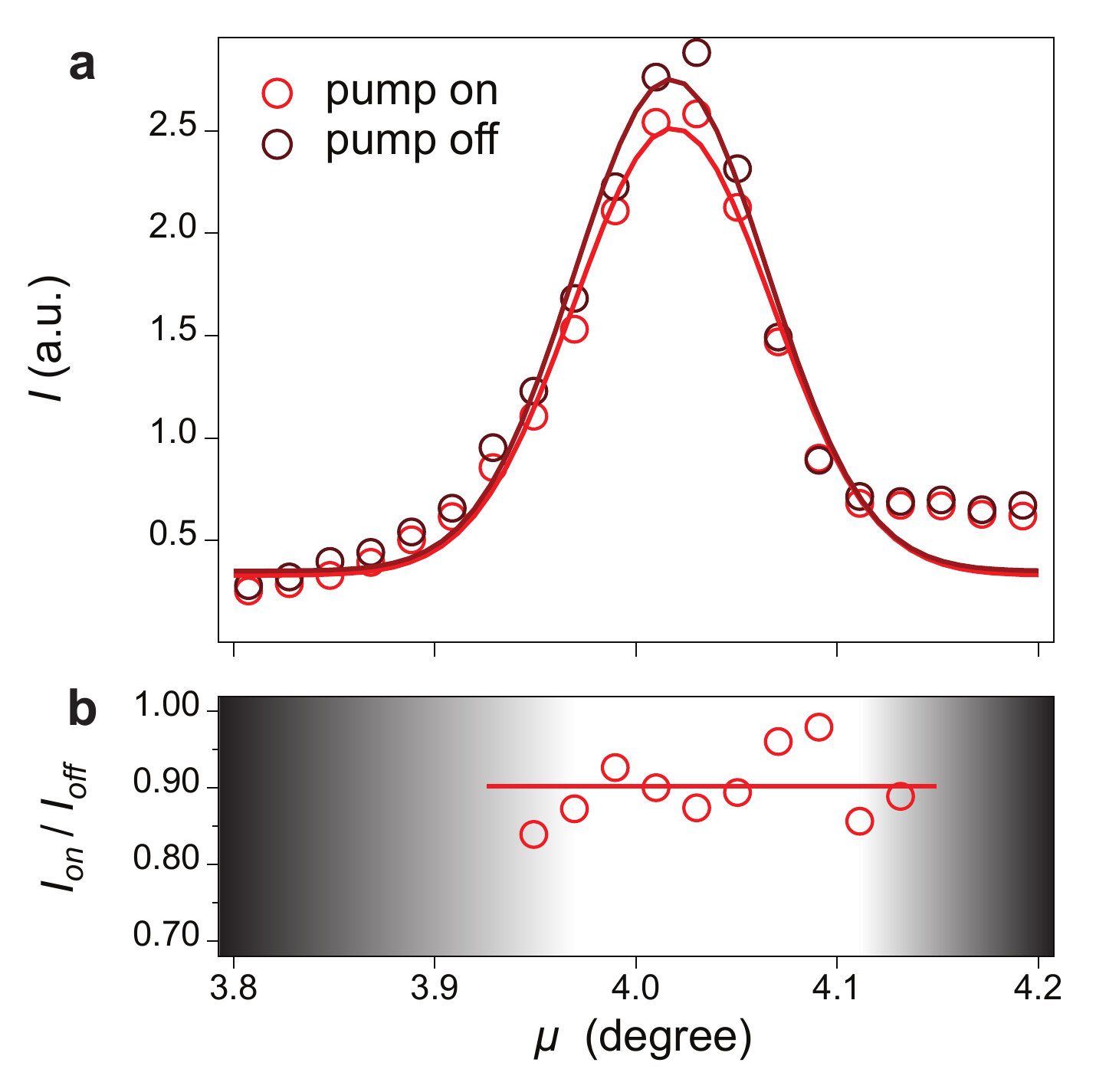}
\label{FigS3}
\caption{\textbf{Rocking curves for the CDW peak at (0 -1.5 2.5) in the presence and absence of pumping.} \textbf{a}, Rocking curves taken at $t=0.5$ ps with (light red) and without (dark red) a pump at $F=$ 1 mJ/cm$^2$. The solid lines are fits to Gaussians. \textbf{b}, Relative light-induced intensity change, which is nearly a constant around the peak region as shown by the solid line.} 
\end{figure}

\section*{S4. Momentum dependence of the CDW peak dynamics}

Previous trXRD studies have reported various possible momentum-($q$) dependent dynamics of CDW. For example, in the cuprate superconductor La$_{2-x}$Ba$_x$CuO$_4$, the recovery time of the intensity close to the CDW superlattice peak exhibits a $q^2$-dependence, indicating diffusive CDW dynamics \cite{AbbamonteSciAdv2019}. Meanwhile, in the case of a rare-earth tritelluride, $q$-dependent dynamics reflecting the formation of CDW domain walls have been observed \cite{TrigoPRB2021}. To test whether the aforementioned physics occurs, we analyze the $q$-dependence of the CDW peak dynamics. As shown in Fig. S4, we find that along both $q_{//}$ and $q_{\perp}$ directions, the CDW melting and recovery dynamics of the peak at (0 -1.5 2.5) do not show noticeable $q$-dependence upon a pump at $F=2.5$ mJ/cm$^2$. The $q$-independence of the CDW dynamics indicates that neither diffusive behavior nor ultrafast domain wall formation is related to our system. However, we do notice that the magnitude of suppression is more significant at larger $q$ away from the CDW diffraction peak, suggesting that the peak width is reduced upon photo-excitation. It is also worthy noting that this phenomenon is more apparent along $q_{\perp}$ than $q_{//}$, demonstrating a more significant decrease of width in the out-of-plane direction, in agreement with the conclusion we draw in the main text. 

\begin{figure*}[th!]
\includegraphics[width=0.8\textwidth]{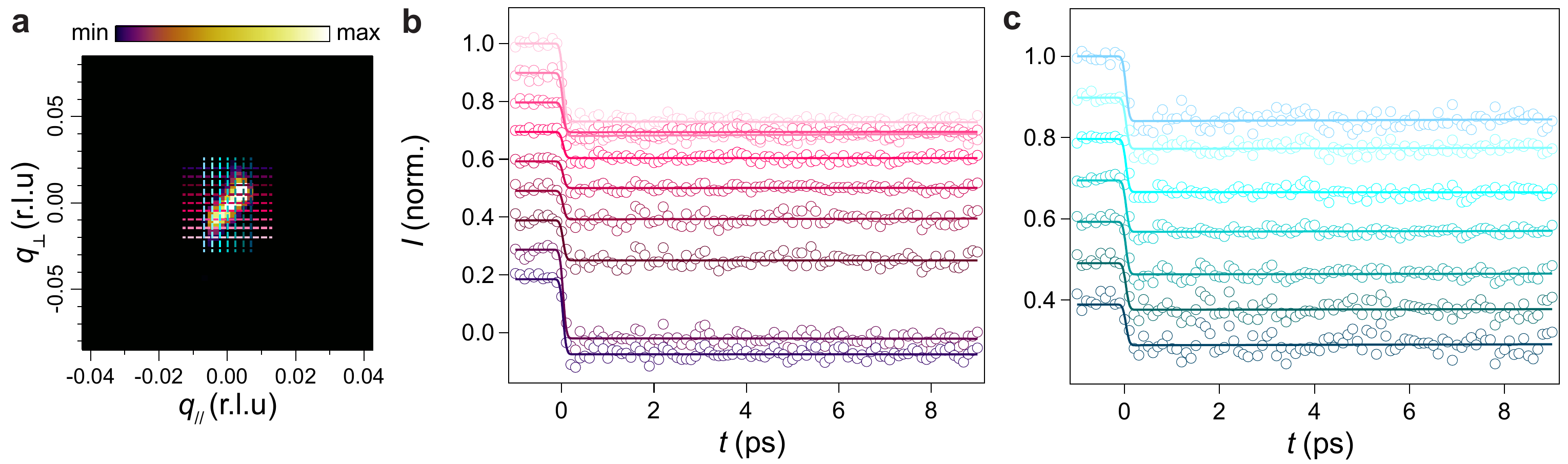}
\label{FigS4}
\caption{\textbf{Temporal evolution of different momentum cuts of the CDW peak at (0 -1.5 2.5).} \textbf{a}, Raw image of the (0 -1.5 2.5) superlattice peak before pumping. Colorscale indicates corresponding counts on the detector. The red and blue dashed lines denote the momentum space cuts along the $q_{//}$ and $q_{\perp}$ directions as shown in panels \textbf{b} and \textbf{c}, respectively. \textbf{b}, Temporal evolution of different $q_{\perp}$ cuts around the CDW peak at (0 -1.5 2.5) acquired at $F=2.5$ mJ/cm$^2$. \textbf{c}, Temporal evolution of different $q_{//}$ cuts around the CDW peak at (0 -1.5 2.5) acquired at $F=2.5$ mJ/cm$^2$. Solid lines are fits to a single-exponential decay. Curves are displaced vertically for clarity.} 
\end{figure*}

\section*{S5. Fluence dependence of the coherent oscillation}

We closely examine the oscillation that appears exclusively in the peak at (0 -1.5 2.5) as a function of pump fluence $F$. The oscillatory components of the peak intensity time traces at various $F$ are depicted in Fig. S5a, and their fast Fourier transform (FFT) spectra are shown in Fig. S5b. A single peak at around 1.3 THz can be seen at all $F$, while at the lowest fluence, another peak may emerge at around 3.1 THz. Both modes have been detected before in transient reflectivity and polarization rotation measurements at lower fluences \cite{HarterPRM2021,WNLPRB2021,WNLPRB2022}. 

We focus on the 1.3 THz mode since it shows up in every trace. We fit the traces shown in Fig. S5a with a single-damped sinusoid $A\exp(-t/\tau)\sin(2\pi ft+\phi)$, where $A$, $\tau$, $f$, and $\phi$ are the amplitude, lifetime, frequency, and phase of the mode, respectively. As shown in Fig. S5c, $f$ does not exhibit softening despite the CDW melting, consistent with the absence of phonon softening observed previously by inelastic X-ray scattering and Raman spectroscopy \cite{MHPRX2021,BlumbergPRB2022,XXXNCOMM2022}. On the other hand, $A$ shows an initial increase with $F$ and then saturates at around $F=0.5$ mJ/cm$^2$ (Fig. S5d). We note that a data point representing zero amplitude at zero pump fluence is included in the plot to depict the trend clearly. 

We would like to comment on the nature of this mode. Previous optical measurements demonstrate that this mode exclusively emerge below $T_c$, suggesting its direct connection to the CDW. Since its frequency matches the frequency of a Cs motion calculated by density functional theory (DFT), the authors assign it to the zone-folded coherent phonon involving Cs motion \cite{HarterPRM2021}. However, previous DFT theories and X-ray diffraction measurements also demonstrate a minor effect of Cs motion on the formation of CDW \cite{WilsonPRM2023,FernandesARXIV2022}. From this perspective, it is less likely that a coherent Cs motion strongly modulates the CDW superlattice peak which arises predominantly from V displacements. On the other hand, the frequency of this mode is very close to the lowest acoustic phonon energy at the CDW reciprocal lattice vector before zone-folding and shows no softening upon temperature changes \cite{MHPRX2021,BlumbergPRB2022,XXXNCOMM2022}. Therefore, it is also possible that this mode is the amplitude mode of $MLL$, which does not soften, signifying its unconventional nature. We note that this does not contradict the recent angle-resolved photoemission results, where the amplitude modes of the $LLL$ phase are found and indeed exhibit softening with $F$ \cite{GedikARXIV2023}. Although our results cannot conclusively determine the origin of this mode, further investigation is required to understand its unusual behavior. Nevertheless, regardless of its nature, our main conclusion remains unchanged. Without loss of generality, in the time-dependent Landau theory (TDLT) simulations in Section 7, we treat this mode as a linearly coupled phonon.

\begin{figure*}[th!]
\includegraphics[width=0.85\textwidth]{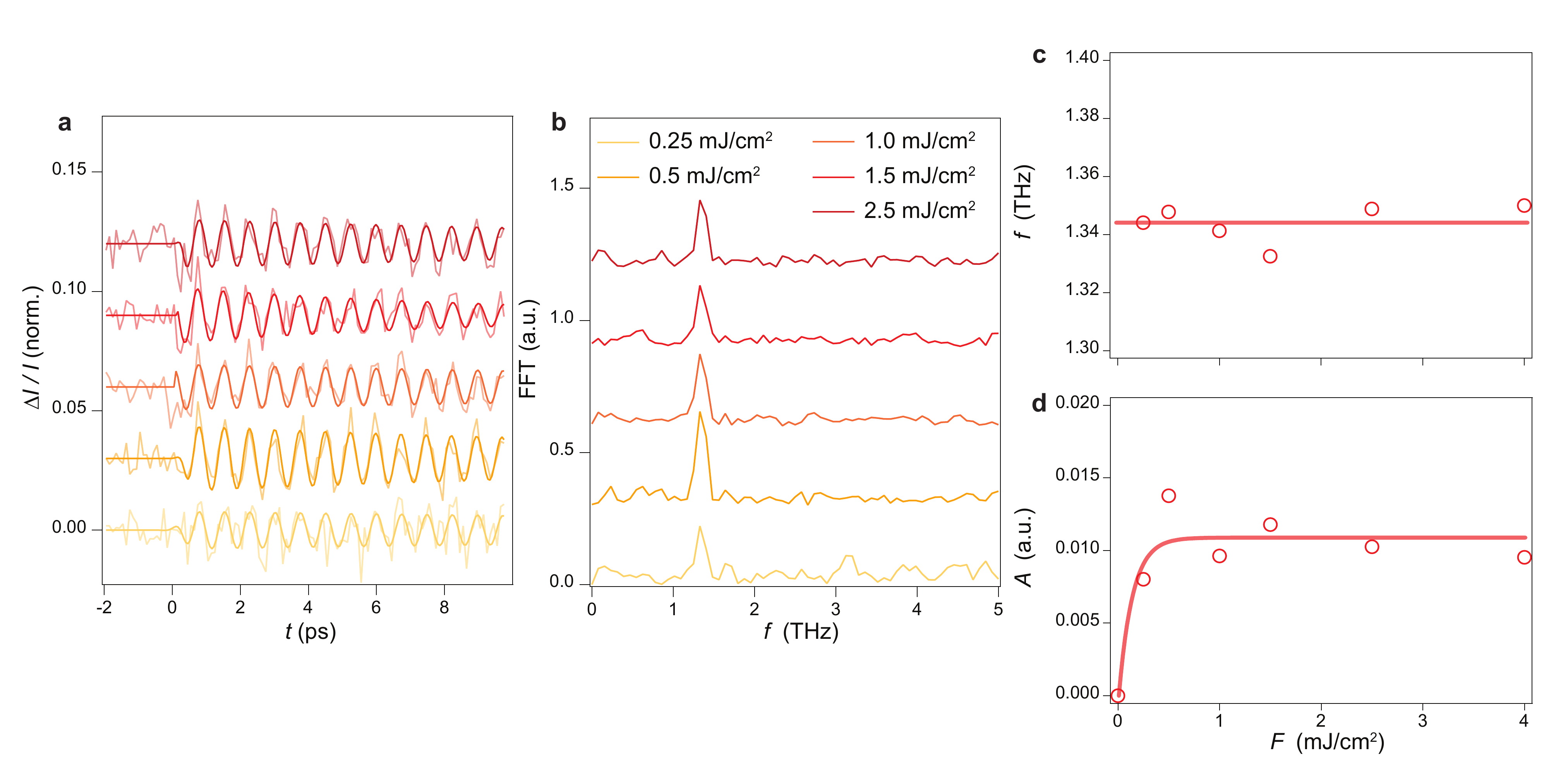}
\label{FigS5}
\caption{\textbf{Fluence dependence of the coherent oscillation.} \textbf{a}, Temporal evolution of the integrated intensity of the peak at (0 -1.5 2.5) subtracted by the single exponential decay background at various pump fluences. Fits to damped sinusoids are overlaid atop the data. Curves are offset for clarity. \textbf{b}, Fourier transform spectra of the curves in panel \textbf{a}. Curves are offset for clarity. \textbf{c}, Fluence dependence of the coherent oscillation frequency. \textbf{d}, Fluence dependence of the coherent oscillation amplitude. Solid lines are guides to eyes. A data point representing zero amplitude at zero pump fluence is included to depict the trend clearly.} 
\end{figure*}

\section*{S6. Temporal evolution of the structural Bragg peaks}

The temporal evolution of two Bragg peaks at (0 0 1) and (0 -1 3) upon photo-excitation is shown in Fig. S5. By integrating the pixels where the Bragg peak is located and subtracting the background, we can obtain the peak intensity change $\Delta I$ as a function of time delay $t$. We can also fit the integrated linecut of the two peaks with a Gaussian along both $q_{//}$ and $q_{\perp}$ directions to obtain their full width at half maximum (FWHM) $\sigma_{//}$ and $\sigma_{\perp}$ as a function of $t$. In the short timescale $t<10$ ps, we do not observe any prominent change in peak intensity and peak width within our signal-to-noise ratio (Fig. S6), in contrast to the melting and sharpening (broadening) of CDW peaks. We indeed observe that the intensity of the peak at (0 0 1) exhibits a slow decrease in the long timescale spanning several hundreds of picoseconds under $F=6$ mJ/cm$^2$, which can be understood as the Debye-Waller effect arising from heating. However, this effect only appears at a relatively long timescale and at fluences higher than the range where we focus on. Therefore, we can conclude that all the transient changes in peak intensity and width of the CDW peaks are nonthermal.

\begin{figure*}[th!]
\includegraphics[width=0.85\textwidth]{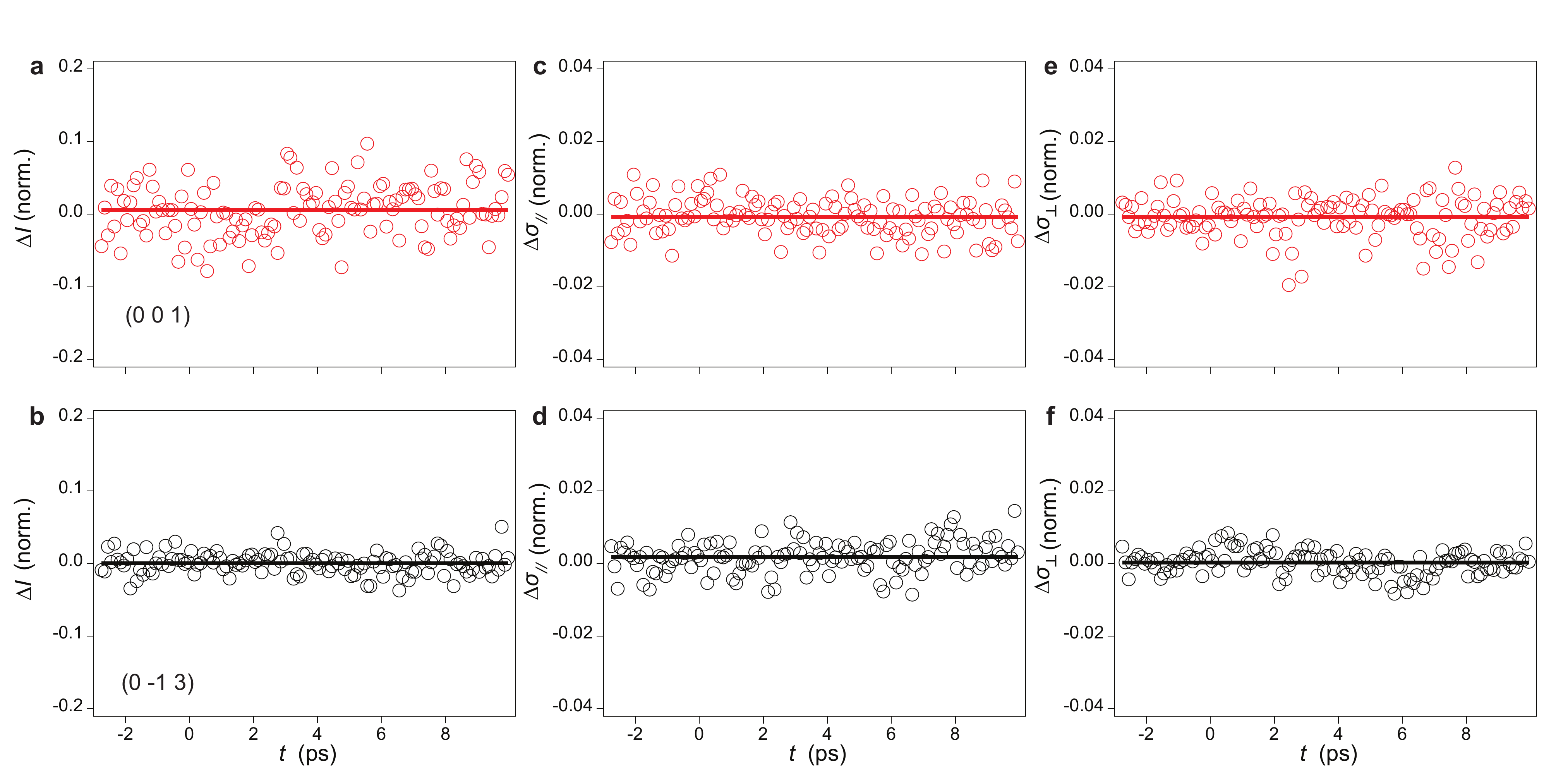}
\label{FigS6}
\caption{\textbf{Temporal evolution of the intensity and width of structural Bragg peaks} \textbf{a,b}, Temporal evolution of the normalized differential intensity of the peak at (0 0 1) and (0 -1 3), respectively. The results are acquired at 30 K with a pump at $F=2.5$ mJ/cm$^2$. \textbf{c, d}, Temporal evolution of the normalized differential FWHM along the in-plane direction of the peak at (0 0 1) and at (0 -1 3), respectively.  \textbf{e, f}, Temporal evolution of the normalized differential FWHM along the out-of-plane direction of the peak at (0 0 1) and at (0 -1 3), respectively. The solid lines in all the panels are guides to eyes.} 
\end{figure*}

\section*{S7. Time-dependent Landau theory simulations}
\subsection*{Static Landau theory}
Previous density functional theory (DFT) calculations have identified two unstable phonon modes in CsV$_3$Sb$_5$, which transform as the $M_1^+$ and $L_2^-$ irreducible representations of the space group P6$/$mmm \cite{HarterPRM2021,FernandesPRB2021,FernandesPRB2022}. These phonon modes occur at three symmetry-equivalent reciprocal lattice vectors located at the three distinct $M$ and three distinct $L$ points in the hexagonal Brillouin zone. The vectors of $M$ are:
\begin{equation}\label{QM}
    Q_M^{(1)}=(0.5\;0\; 0), \;Q_M^{(2)}=(0\;0.5\;0), \;Q_M^{(3)}=(-0.5\;-0.5\;0),
\end{equation}
and those of $L$ are:
\begin{equation}\label{QL}
    Q_L^{(1)}=(0.5\;0\; 0.5), \;Q_L^{(2)}=(0\;0.5\;0.5), \;Q_L^{(3)}=(-0.5\;-0.5\;0.5),
\end{equation}
In real space, both $M_i$ and $L_i$ modes primarily involve nearest neighbour V-V bond contraction as shown in Fig. 1 in the main text, where the in-plane displacements of the V atoms account for over 90\% of the total CDW displacements \cite{FernandesPRB2021}. The distinction between $M$ and $L$ is in the relative phase of the displacements in consecutive kagome layers along the $c$-axis: the V atoms are displaced in-phase between the neighbouring layers for the $M$ mode, while the displacement is out-of-phase between the neighbouring layers for the $L$ mode. Clearly, each $M_i$ forms a stripe pattern that doubles the unit cell along the stripe direction, while for $L_i$, the unit cell is also doubled along the $c$-axis.

The different ``3Q" CDW structures correspond to distinct superpositions of three $M_i$ and $L_i$ CDW order parameters. We denote the phases formed by $M$ and $L$ modes using a general notation $M_1M_2M_3+L_1L_2L_3$. A distortion with the opposite sign, namely an expansion of the V-V bond, is denoted as $\overline{M}$ or $\overline{L}$. For distortions where all the $M$ or $L$ order parameters have zero amplitude, we simplify the notation to $L_1L_2L_3$ and $M_1M_2M_3$, respectively. Also, we note that domains equivalent to each other up to a translation or rotation can be obtained as long as the sign of production $M_1M_2M_3$ and $L_1L_2L_3$ is preserved. For instance, $M_1M_2M_3=\overline{M_1}\overline{M_2}M_3$ and $L_1\overline{L_2}L_3=\overline{L_1L_2L_3}$.

The Landau free energy functional for the coupled $M_i$ and $L_i$ order parameters is determined by symmetry, and its general form expanded up to the quartic order can be expressed as:
\begin{equation}\label{FreeEnergy}
    F=F_M+F_L+F_{ML},
\end{equation}
where
\begin{equation}
\label{FreeEnergyDifferentTerms}
\begin{split}
    F_M=&\frac{\alpha_M(1-T/T_M)}{2}M^2+\frac{\beta_M}{3}M_1M_2M_3+\frac{u_M}{4}M^4+\\&\frac{\lambda_M}{4}(M_1^2M_2^2+M_2^2M_3^2+M_1^2M_3^2),\\
    F_L=&\frac{\alpha_L(1-T/T_L)}{2}L^2+\frac{u_L}{4}L^4+\\&\frac{\lambda_L}{4}(L_1^2L_2^2+L_2^2L_3^2+L_1^2L_3^2),\\
     F_{ML} =&\frac{\beta_{ML}}{3}(M_1L_2L_3+L_1M_2L_3+L_1L_2M_3)+\\&\frac{\lambda_{ML}^{(1)}}{4}(M_1M_2L_1L_2+M_2M_3L_2L_3+M_1M_3L_1L_3)+\\
    &\frac{\lambda_{ML}^{(2)}}{4}(M_1^2L_1^2+M_2^2L_2^2+M_3^2L_3^2)+\frac{\lambda_{ML}^{(3)}}{4}M^2L^2,
\end{split}
\end{equation}
where $M^2=M_1^2+M_2^2+M_3^2$ and $M^4=(M^2)^2$, with $L^2$ and $L^4$ defined analogously. $\alpha_M$ and $\alpha_L$ are the coefficients of the quadratic terms with tunable temperature $T$. Also note that since $M_i$ and $L_i$ belong to different irreducible representations, $\alpha_M$ and $\alpha_L$ are not necessarily identical, manifested as different transition temperatures $T_M$ and $T_L$ in the absence of coupling. Both $\alpha_M$ and $\alpha_L$ should be negative so that the ordered phase can be realized when $T<T_{M/L}$. $u_M$, $u_L$, $\lambda_M$, $\lambda_L$, $\lambda_{ML}^{(1)}$, $\lambda_{ML}^{(2)}$, and $\lambda_{ML}^{(3)}$ are the coefficients of the quartic terms. $\beta_M$ and $\beta_{ML}$ are the coefficients of the cubic terms.

One unambiguous feature of the functional is that while $F_M$ and $F_{ML}$ contain a trilinear term, $F_L$ does not. This arises from the fact that different reciprocal lattice vectors of $M_i$ and $L_i$ obey different symmetries: $Q_M^{(1)}+Q_M^{(2)}+Q_M^{(3)}$ and $Q_M^{(1)}+Q_L^{(2)}+Q_L^{(3)}$ are zero modulo a reciprocal lattice vector, and thus the trilinear term is allowed. On the other hand, $Q_L^{(1)}+Q_L^{(2)}+Q_L^{(3)}\neq0$, and thus the trilinear term $L_1L_2L_3$ is prohibited. Interestingly, this difference renders M more of a first-order nature, while L is more of second-order, as can be seen from $F_M$ and $F_L$. This is also manifested in the different temperature dependence of the CDW superlattice peak intensity, where the peaks contributed exclusively by $L$ exhibit a smoother increase at the transition while the peaks contributed by both show a more abrupt increase due to the first-order characteristic of $M$ (Fig. S2). Additionally, the presence of the trilinear term leads to $M_i\neq\overline{M_i}$, but its absence yields $L_i=\overline{L_i}$. One can intuitively understand this by examining the real-space CDW superlattice structure: $M_1M_2M_3(+000)$ corresponds to a structure in which the two neighbouring kagome planes form aligned inverse star-of-David (soD) distortions, while $\overline{M_1}M_2M_3$ corresponds to a distortion where the two neighbouring kagome planes form aligned soD patterns. These two superlattice structures are not necessarily energy-equivalent. However, both $(000+)L_1L_2L_3$ and $\overline{L_1}L_2L_3$ correspond to different domains of the same superlattice with aligned alternative inverse soD and soD planes and are identical up to a translation along $c$-axis.

A combination of instabilities at $M$ and $L$ points finally leads to six possible $2\cross2\cross2$ CDW structures, namely $\overline{M_1}M_2M_3$ (aligned soD), $M_1M_2M_3$ (aligned inverse soD), $L_1L_2L_3$ (aligned alternating inverse soD and soD), $\overline{M_1}00+0L_2L_3$ (staggered soD), $M_100+0L_2L_3$ (staggered inverse soD), and $M_1M_20+00L_3$ (staggered alternating inverse soD and soD) \cite{CominNMAT2023}. Intriguingly, these states are not always thermodynamically stable in the presence of the trilinear coupling term $\beta_{ML}$. For instance, the $L_1L_2L_3$ phase is inevitably accompanied by a finite amount of $M_1M_2M_3$ or $\overline{M_1}M_2M_3$ and becomes the $L_1L_2L_3+M_1M_2M_3$ intertwined phase \cite{FernandesPRB2021}, which has been confirmed by the recent angle-resolved photoemission spectroscopy results \cite{CominNMAT2023}. Similarly, the $M_1M_20+00L_3$ phase acquires additional $M$ and $L$ distortions from the two trilinear terms and cannot emerge as the leading instability \cite{FernandesPRB2021}.

All the Landau coefficients have been obtained from recent DFT calculations \cite{FernandesARXIV2022}: $\alpha_M=-2.53$ eV/\AA$^2$, $\alpha_L=-2.53$ eV/\AA$^2$, $\beta_M=-22.16$ eV/\AA$^3$, $\beta_{ML}=-24.15$ eV/\AA$^3$, $u_M=76.43$ eV/\AA$^4$, $u_L=89.93$ eV/\AA$^4$, $\lambda_M=-137.67$ eV/\AA$^4$, $\lambda_L=-194.47$ eV/\AA$^4$, $\lambda_{ML}^{(1)}=347.73$ eV/\AA$^4$, $\lambda_{ML}^{(2)}=332.28$ eV/\AA$^4$, $\lambda_{ML}^{(3)}=24.20$ eV/\AA$^4$. Based on these parameters, free energy landscape of different CDW structures can be simulated and a minimization of the free energy yields the values of the order parameters. 

At zero temperature, $M_100+0L_2L_3$ is found to have the lowest energy, in agreement with a wide array of experimental observations \cite{PYYPRR2023,GeckPRB2022,WilsonPRM2023,WLNPHY2022,WNLPRB2022,CXHNature2022}. At elevated temperatures, a reasonable choice of $T_L$ and $T_M$ ($T_L$=91 K and $T_M=85$ K$<T_L$) predicts a ground state of $L_1L_2L_3+M_1M_2M_3$, in line with another set of observations \cite{SMPRB2022,TjernbergPRR2022,CominNMAT2023,CanosaNCOMM2023}. The Landau theory simulations thus reproduce the two experimentally proposed CDW structures and demonstrate that they are energetically similar. For simplicity, we denote these two phases as $MLL$ and $LLL$ in the following discussions ($MMM$ notation is dropped for $LLL+MMM$ because the $M$ distortion is much smaller than the $L$ distortion as shown shortly after, but the non-zero $M$ is preserved). The Landau free energy for the two phases can be thus simplified as:
\begin{equation}\label{FreeEnergyMLL}
\begin{split}
    F_{MLL}=&\frac{\alpha_M(1-T/T_M)}{2}M^2+\frac{u_M}{4}M^4+\\
    &\frac{2\alpha_L(1-T/T_L)}{2}L^2+\frac{4u_L+\lambda_L}{4}L^4+\\
    &\frac{\beta_{ML}}{3}ML^2+\frac{2\lambda_{ML}^{(3)}}{4}M^2L^2,
\end{split}
\end{equation}
\begin{equation}\label{FreeEnergyLLL}
\begin{split}
    F_{LLL}=&\frac{3\alpha_M(1-T/T_M)}{2}M^2+\frac{\beta_M}{3}M^3+\frac{9u_M+3\lambda_M}{4}M^4+\\
    &\frac{3\alpha_L(1-T/T_L)}{2}L^2+\frac{9u_L+3\lambda_L}{4}L^4+\\
    &\frac{3\beta_{ML}}{3}ML^2+\frac{3\lambda_{ML}^{(1)}+3\lambda_{ML}^{(2)}+9\lambda_{ML}^{(3)}}{4}M^2L^2,
\end{split}
\end{equation}
where for $MLL$ we set $M_1=M$, $L_2=L_3=L$, and for $LLL$, $M_1=M_2=M_3=M$,$L_1=L_2=L_3=L$. 

The free energy landscapes of the aforementioned two phases are shown in Fig. S7a and Fig. S8a, respectively. Both potentials unambiguously depict an asymmetry along the $M$ coordinate and a symmetry along the $L$ coordinate due to the trilinear coupling terms, in agreement with our predictions. The $MLL$ free energy potential shows two equivalent minima at both finite $M$ and $L$ and their distortion amplitudes are also nearly identical, i.e. $M\sim L$. This confirms that a staggered inverse soD ground state is energetically favorable. On the other hand, for the $LLL$ free energy potential, the distortion amplitude of $M$ is significantly smaller than that of $L$ when the energy is minimized, in line with the angle-resolved photoemission spectroscopy results, where the former is 7 times weaker than the latter \cite{CominNMAT2023}. Also note that the coexistence of finite $M$ and $L$ indicates that they form intertwined order and are spatially homogeneous. The resulting structure is still an aligned alternating inverse soD and soD, but the in-plane distortion value is different for the two kagome planes ($L-M$ and $L+M$, respectively).

\subsection*{Time-dependent Landau theory simulations}

We now derive the equations of motion for the order parameters. For the phenomenological treatment, whether $M_i$ and $L_i$ are lattice distortions or electronic order parameters of CDW is not critical, since they transform as the same irreducible representations. We assume they are electronic and thus display overdamped dynamics since on the phononic timescale, electrons nearly adiabatically adjust themselves to the local minimum of the Landau potential \cite{DemsarPRL2010,GedikPRL2019}. The simulation results will not qualitatively change if we assume they are phononic and display second-order dynamics. To account for the phonon oscillation that exclusively occurs in L\:=\:half integer peaks, we introduce a lattice order parameter $X$ in a harmonic potential, which primarily linearly couples to $L$ and displays underdamped (second-order) dynamics \cite{DemsarPRL2010,GedikPRL2019}:
\begin{equation}\label{FreeEnergywithLattice}
    F=F_{MLL/LLL}+F_X,
\end{equation}
where
\begin{equation}\label{FreeEnergywithLattice}
    F_X=\frac{1}{2}\omega^2X^2+gXL,
\end{equation}
where $\omega$ is the phonon frequency ($2\pi\cross1.3$ THz) and $g$ is the electron-phonon coupling constant. \textcolor{black}{Our approach by linearly coupling a phonon to $L$ serves the exactly identical purpose as adding a phonon which only affects L=half integer peaks. The reason why we did not adopt the latter is because M and L involve V atom displacements, while previous literature has demonstrated the phonon we observed involves Cs atom motion (Section 5). Based on the principle of parsimony, we model the Cs motion as a linearly coupled phonon to L instead of involving additional atoms in the structure factor calculation.}

We also assume that the photo-excitation acts as a homogeneous energy quench to the quadratic term $\alpha_{M/L}$, which is proportional to the pump fluence, while the other Landau parameters are independent on time:

\begin{equation}\label{FreeEnergywithLattice}
    \alpha_{M/L}(t)=\alpha_{M/L}(1-T/T_{M/L})[1-\kappa_{M/L}F\frac{(1+\erf{\frac{t}{\sqrt{2}t_r}})}{2}(e^{-t/\tau}+C)].
\end{equation}

Here, $F$ is pump fluence, $\kappa_{M/L}$ is a proportionality constant for normalizing $F$ but they can be different for $M$ and $L$, $\erf$ is the error function, $t_r$ is the rise time of the quasiparticles after excitation, $\tau$ is the thermal relaxation time of the quasiparticles, and $C$ is the long-term background constant within the simulation time window. The form of the time-dependent quench mimics the carrier excitation and relaxation dynamics measured by transient reflectivity experiments \cite{HarterPRM2021, WNLPRB2021}.

Based on these assumptions, the dynamical equations of order parameters can be expressed as:
\begin{equation}\label{Dynamicalequ}
\begin{split}
    \frac{1}{\gamma_{M}}{{\partial}_t}{M}&=-\frac{\partial F}{\partial M},\\
    \frac{1}{\gamma_{L}}{{\partial}_t}{L}&=-\frac{\partial F}{\partial L},\\
    {{\partial}_t}^2{X}=-&2\gamma_X{\partial}_t{X}-\frac{\partial F}{\partial X},
\end{split}
\end{equation}
where $\gamma_M$, $\gamma_L$, and $\gamma_X$ are phenomenological decay constants to account for damping of electronic and structural order parameters. 

We first investigate the dynamics of the $MLL$ phase. We assume $t_r=0.1$ ps, $\tau=1$ ps, $C=0.9$, $\gamma_M=1$ THz, $\gamma_L=0.5$ THz, $\gamma_X=0.2$ THz. The initial values of $M$, $L$, and $X$ are set at their local minima. We also carefully choose the fluence regime and tentatively set $\kappa_M=1$ and $\kappa_L=3$, i.e. $L$ is roughly three times more susceptible to light excitation than $M$ to best reproduce the experimental results. The simulated temporal evolution of $L$ and $M$ normalized to their equilibrium values are shown in Figs. S7b and S7c, respectively. It is evident that neither of the two order parameters is fully quenched in this fluence regime, but $L$ indeed shows a more substantial suppression than $M$. Additionally, the linearly coupled phonon exhibits a pronounced oscillation in $L$ because of the linear coupling. However, due to the coupling between $M$ and $L$, the oscillation also exists in $M$, albeit weakly. 

Using the temporal evolution of $M$ and $L$ obtained from the simulation, we can calculate the dynamics of peak intensity $I(t)$ by incorporating their dynamics into the SF of different peaks: $\delta x_i(t)$ and $\delta y_i(t)$ can be expressed as a linear superposition of $M$ and $L$. The simulated dynamics of each peak are shown in Fig. S7d-e. The peak at (0 -1.5 2.5) shows a considerable decrease accompanied by phonon oscillation upon light excitation (Fig. S7d). This L\:=\:half-integer peak mainly reflects the dynamics of the $L$ order parameter, as $M$ can not generate L\:=\:half-integer peaks. In contrast, the peak at (-0.5 -1 2) exhibits a smaller intensity drop with faint oscillation, while the peak at (0 -1.5 3) exhibits a drastic drop similar to the peak at (0 -1.5 2.5). 

To emulate the measured intensity change, the dramatic penetration depth mismatch between the pump and the probe pulses at different wavelength ranges should be considered. Due to the pump-probe penetration depth mismatch, different depths $z$ below the sample surface experience a different level of quenching of the potential. The reported pump and probe penetration depths are $\delta_{pu}=$78 nm \cite{WHHPRB2021} and $\delta_{pr}=$568 nm, respectively, which is approximately a seven-fold difference. To simulate the effects of the penetration depth mismatch, we assume the sample is composed of thin layers with thickness $d=$ 1 nm and an exponential $z$-dependence of quenching with $F(z)=F\exp(-z/\delta_{pu})$ arising from an exponentially decaying photo-carrier distribution. We then solve the temporal evolution of $M(z,t)$, $L(z,t)$, and $I(z,t)$ for each layer. The probed-region-integrated value of $I_{int}(t)$ is calculated by summing over all the layers with each layer weighted by the probe penetration depth: $I_{int}(t)=\sum^\infty_{z=0}\exp(-z/\delta_{pr})I(z,t)$. The integrated simulation results are shown in Figs. 2d and 2e. Figs. S7d-f can thus be understood as showing the dynamics of peak intensity from the top excited layer. A comparison between $I_{int}$ and $I$ reveals a 100\% reduction of $I$ should correspond to a nearly 18\% drop in $I_{int}$ (Fig. S13).

Although at first glance, it is plausible to attribute all three peaks to the $MLL$ phase, two features suggest a different origin of the peak at (0 -1.5 3). First, the intensity drop observed in peaks (0 -1.5 2.5) and (0 -1.5 3) are very similar. \textcolor{black}{After considering penetration depth mismatch, they are still quantitatively similar (Fig. S7g-i), whereas} the experimental results show that the reduction of the latter is more considerably larger than the former. Second, since both order parameters are far from complete suppression, the simulated dynamics of all the peaks exhibit no clear features of dynamical slowing down of recovery time \cite{GedikPRL2019}, at odds with the experimental observation of an apparent slowing down of dynamics at peak (0 -1.5 3).

\begin{figure*}[th!]
\centering
\includegraphics[width=0.85\textwidth]{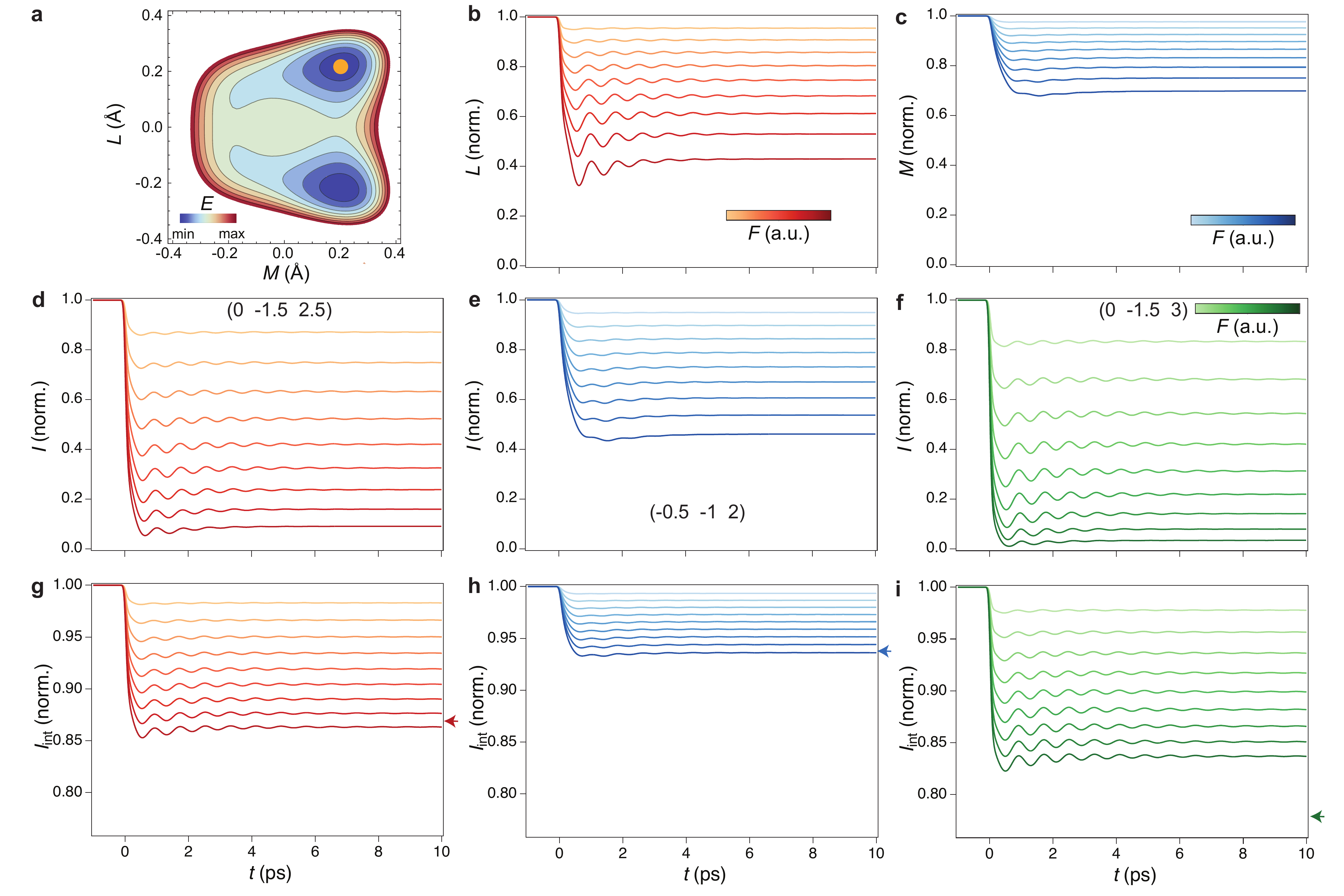}
\label{FigS7}
\caption{\textbf{TDLT simulations of the $MLL$ phase.} \textbf{a}, Free energy landscape of the $MLL$ phase. The initial ground state is marked by a yellow dot. \textbf{b,c}, Temporal evolution of $L$ and $M$ order parameters at various fluences, respectively. \textbf{d-f}, Temporal evolution of intensity change of peak at (0 -1.5 2.5), (-0.5 -1 2), and (0 -1.5 3) at various pump fluences. \textcolor{black}{\textbf{g-i}, Temporal evolution of the integrated intensity change of peak at (0 -1.5 2.5), (-0.5 -1 2), and (0 -1.5 3) at various pump fluences. The maximal intensity decrease measured experimentally are marked by arrows.} The fluence range is identical in all the panels.} 
\end{figure*}

\begin{figure*}[th!]
\includegraphics[width=6.75in]{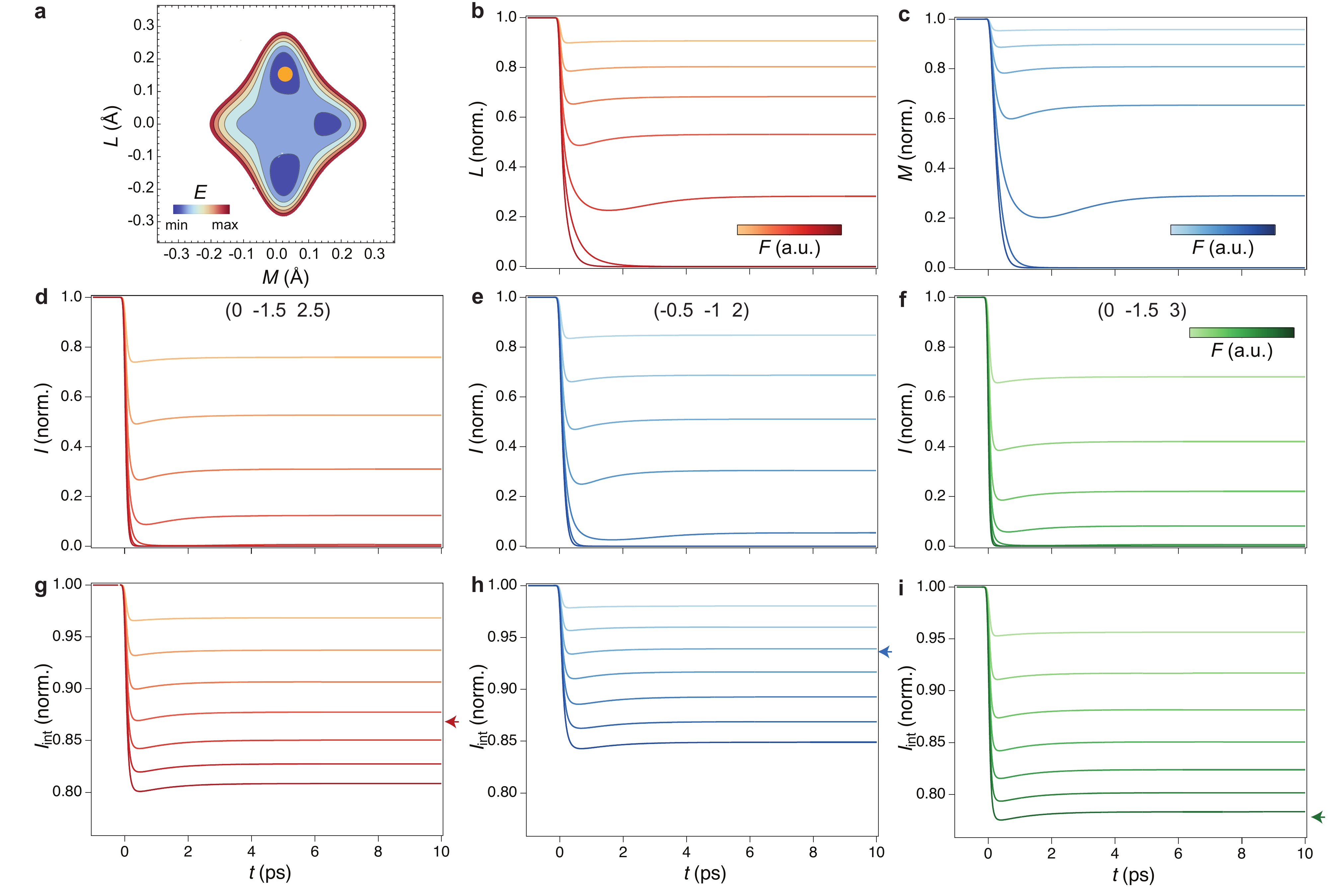}
\label{FigS8}
\caption{\textbf{TDLT simulations of the $LLL$ phase.} \textbf{a}, Free energy landscape of the $LLL$ phase. The inital ground state is marked by a yellow dot. \textbf{b,c}, Temporal evolution of $L$ and $M$ order parameters at various fluences, respectively. \textbf{d-f}, Temporal evolution of intensity change of peak at (0 -1.5 2.5), (-0.5 -1 2), and (0 -1.5 3) at various pump fluences. \textcolor{black}{\textbf{g-i}, Temporal evolution of the integrated intensity change of peak at (0 -1.5 2.5), (-0.5 -1 2), and (0 -1.5 3) at various pump fluences. The maximal intensity decrease measured experimentally are marked by arrows.} The fluence range is identical in all the panels.} 
\end{figure*}

It is thus reasonable to suspect that the peak at (0 -1.5 3) may mainly originate from the $LLL$ phase, since based on the SF calculation in Section 1, both structures can contribute to this peak with similar amplitude. To investigate this, we simulate the dynamics of $M$ and $L$ in the $LLL$ phase (Figs. S8b and S8c). We explore the same fluence range employing the same set of parameters as we have used for $MLL$ except for two: $T_M$, which we set to be $10 K$, or at $T=30$ K $LLL$ will not be the global minimum but a local minimum, and $g$, which we set to be 0, since the peak at (0 -1.5 3) does not exhibit phonon oscillation within our resolution. \textcolor{black}{However, we note that the absence of phonon should not be used as evidence for coexistence of two phases. We set $g=0$ to better match the experimental observations here. Also} note that although we show the normalized intensity change here, the absolute value of $L$ is about 9 times larger than $M$, which matches the experimental value \cite{CominNMAT2023}. 

In contrast to the $MLL$ case, despite the 3-times higher sensitivity of $L$ in response to a light excitation, both $M$ and $L$ in $LLL$ depict a similar level of melting at the same fluence and a complete melting of both is realized at an intermediate fluence. This may arise from the fact that the equilibrium value of $M$ is so small that a moderate light excitation can fully quench it. In addition, a dynamical slowing down of the melting and recovery time can be observed at the critical fluence where both order parameters are completely quenched. Employing the same aforementioned method, we calculate the peak intensity dynamics $I(t)$ at the three peak positions (Fig. S8d-f) and include the pump-probe penetration depth mismatch to emulate the experimental case (Fig.  Sg-i). Again, we note that the absolute equilibrium intensity of the peak at (0 -1.5 2.5) is two orders of magnitude smaller than the other two peaks and the same peak in the $MLL$ phase, confirming that this peak should be predominantly contributed by the $MLL$ phase. Although the absolute equilibrium intensity of the peak at (-0.5 -1 2) is similar to that of (0 -1.5 3) in $LLL$, its intensity is two orders of magnitude smaller than the same peak contributed by the $MLL$ phase. 

Dynamics of $I(t)$ also confirm that the first two peaks cannot mainly arise from $LLL$, because akin to $M$ and $L$, all the peaks show similar levels of intensity drop at the same fluence (Fig. S8d-i), in contrast to the dramatically different experimental behaviors. Moreover, all the peaks exhibit dynamical slowing down around the same fluence, which is absent in the experimental data. However, the behavior of peak (0 -1.5 3) qualitatively agrees with the experimentally measured dynamics, and the intensity drop in integrated $I_{int}(t)$ quantitatively agrees with the experimental value (Fig. S8i), demonstrating that the peak at (0 -1.5 3) mainly reflects the dynamics of the $LLL$ phase.

We would like to note that despite its simplicity, the phenomenological TDLT simulation accurately predicts the dynamics of various peaks and disentangle the order parameter evolution of different phases. With almost no adjustable parameters in our calculation, it is encouraging to see a reasonable match in the absolute value of the CDW suppression between experiments and simulations. Indeed, a few ingredients can be further added to the model to improve its quantitative agreement with the measurements, such as fluence-dependent relaxation time $\tau$, spatial inhomogeneity, and higher-order phonon couplings.

\subsection*{Robustness of the simulation results}

Here, we demonstrate the robustness of the TDLT simulation results against variations of the simulation parameters. The simulation parameters we employed can be categorized into three groups: 1) microscopic parameters $\alpha$’s, $\beta$’s, $u$’s, $\lambda$’s, and $T$’s determining the shape of the potential energy surface (PES) and equilibrium values of the order parameters $M$ and $L$. These values are adopted from previous first-principles calculations \cite{FernandesARXIV2022}. Modifying them is expected to modulate the PES shape, while the peak intensity decrease should remain qualitatively similar with a scaling of the pump fluence. 2) dynamical parameters $\kappa_{L/M}$, $t_r$, $\tau$, $C$, $\gamma_{L/M}$, $\omega$, and $g$, which determine the rise time, decay time, oscillation frequency, oscillation damping time, and oscillation amplitude of the order parameters. Among these parameters, only $\kappa_{L/M}$ determines the photo-susceptibility (the amount of melting upon 1 mJ/cm$^2$ pumping) of $M$ and $L$ and thus the relative intensity drop of different peaks, while the others are less relevant to the magnitude of intensity decrease. 3) penetration depths of pump $\delta_{pu}$ and probe $\delta_{pr}$, which are critical in determining the measured peak intensity decrease.

To assess the impact of parameter variations on peak intensity decreases, we performed more comprehensive TDLT simulations by considerably varying all three groups of parameters. In Fig. S9, we plot the simulated normalized integrated intensity $I_{int}$ for the three targeted peaks with different sets of parameters as a function of the pump fluence $F$ at $t$ = 0.5 ps where the drop is maximal. To compare the relative intensity drops of different peaks between different sets of parameters, we apply a scaling factor for all the peaks along the pump fluence axis. This is valid because there exists a scaling factor between simulated and experimental pump fluence. Fig. S9a demonstrates that halving or doubling all the microscopic parameters minimally affects the relative magnitude of intensity decrease ($\sim$1\%), in line with our expectation. Fig. S9b explores a large range of the ratio $\kappa_{L} / \kappa_{M}$, revealing qualitative consistency in peak intensity decrease of different peaks for $\kappa_{L} / \kappa_{M}$ $\in$ (2, 5), with $\kappa_{L} / \kappa_{M}$ $\sim$ 3 providing the closest agreement with experimental results. This confirms that $L$ is more susceptible than $M$ against photo-excitation and variations in dynamical parameters do not qualitatively alter the simulation results. As detailed in Supplementary Section 13, reasonable choices for penetration depths do not impact the peak dynamics qualitatively. Therefore, we can conclude that our simulation results quantitatively match the experimental values against even considerable variations in all the parameters.

\begin{figure*}[th!]
\centering
\includegraphics[width=1.0\textwidth]{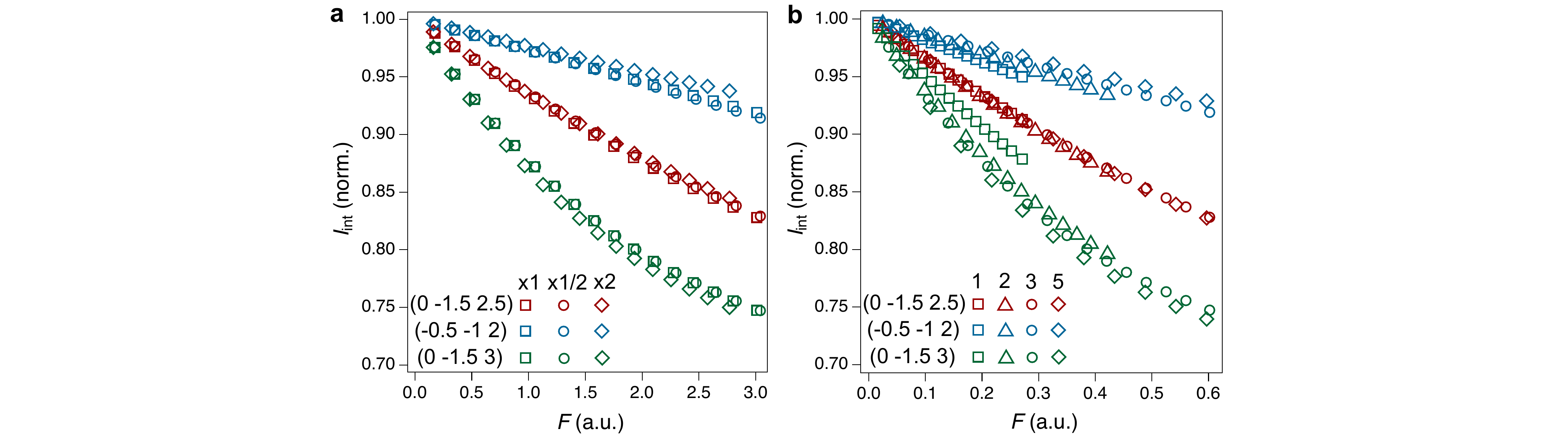}
\label{FigS9}
\caption{\textbf{a}, Simulated fluence dependence of the normalized integrated intensity of the three peaks with all the microscopic parameters ($\alpha$’s, $\beta$’s, $u$’s, and $\lambda$’s) equal to the values obtained from the first-principles calculations ($\cross1$), half of those values ($\cross1/2$), and double of those values ($\cross2$). All the dynamical parameters and the penetration depths are identical to the values that have been used in this section. \textbf{b}, Simulated fluence dependence of the normalized integrated intensity of the three peaks with $\kappa_{L}/\kappa_{M}$ equal to 1, 2, 3, and 5. All the microscopic parameters, the other dynamical parameters, and the penetration depths are identical to the values that have been used in this section.} 
\end{figure*}

\section*{S8. Temporal evolution of the width of the peak at (-0.5 -1 2)}

Temporal evolution of the width of the peak at (-0.5 -1 2) upon photo-excitation exhibits qualitatively similar features to that of the peak located at (0 -1.5 2.5) (Fig. S10). Moreover, the decrease values of both the in-plane ($\sigma_{//}$) and out-of-plane ($\sigma_{\perp}$) directions of (-0.5 -1 2) and (0 -1.5 2.5) show good quantitative agreement. This observation suggests that both peaks possibly arise from the same CDW phase ($MLL$), even though there is a significant difference in the degree of reduction in peak intensity. 

\begin{figure*}[th!]
\centering
\includegraphics[width=0.85\textwidth]{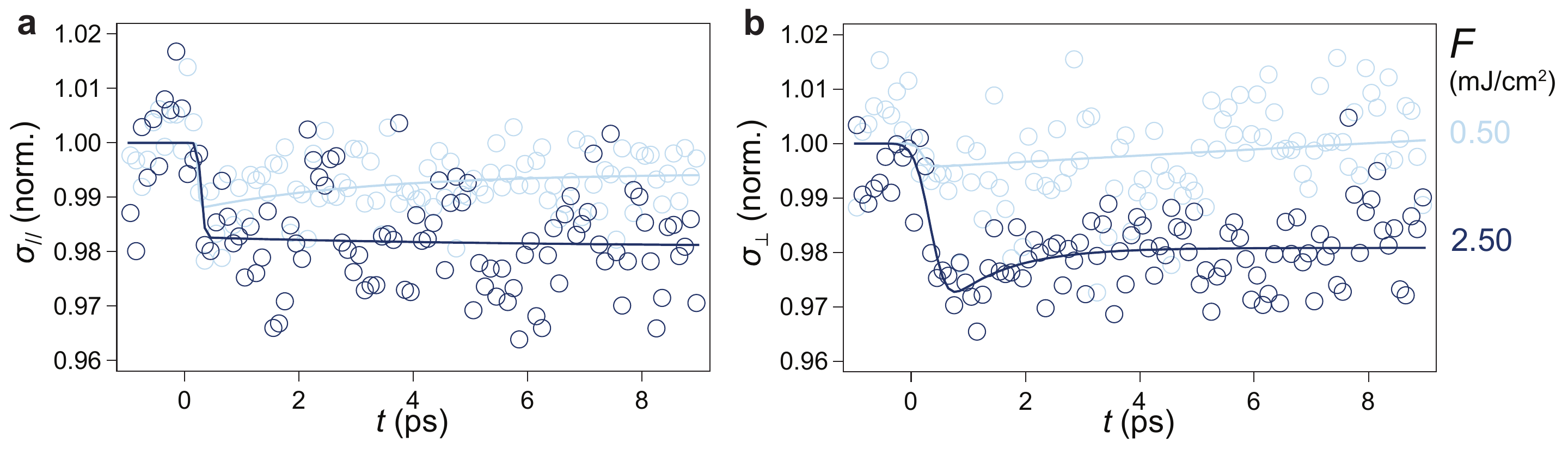}
\label{FigS10}
\caption{\textbf{Temporal evolution of the width of the peak at (-0.5 -1 2)} \textbf{a,b}, Normalized width change of the peak at (-0.5 -1 2) along the in-plane and the out-of-plane directions, respectively, acquired at F = 0.5 and 2.5 mJ/cm2. Solid lines are fits to a single-exponential decay.} 
\end{figure*}

\section*{S9. Temperature dependence of the CDW peak width}

Figure S11 shows that the width of the CDW satellite peaks as a function of temperature around $T_c$ = 91 K in the absence of the pump pulse. As shown by the normalized linecuts along $q_{\perp}$, the peak width increases as the temperature approaches $T_c$ from below (Fig. S11a). We fit the linecuts of different peaks along both $q_{\perp}$ and $q_{//}$ directions with a Gaussian and obtain the half width at the maximum (HWHM) $\sigma_{//}$ and $\sigma_{\perp}$ as functions of temperature (Fig. S11b-e). The width of all the investigated CDW peaks moderately increases when $T<90.5$ K and quickly diverges when $T$ approaches $T_c$, indicating a significant decrease in the correlation length of each CDW domain. This is in sharp contrast to the temporal dynamics of the peak width at (0 -1.5 2.5), where the peak width decreases after pumping. Therefore, we conclude that the anomalous photo-induced sharpening of the peak at (0 -1.5 2.5) is not due to a thermal effect.

Also we note that both $\sigma_{//}$ and $\sigma_{\perp}$ are presented in reciprocal lattice units (r.l.u.). We thus get the correlation length of the $MLL$ phase at around 89 K is around 250 unit cells within the kagome plane and around 80 unit cells across the kagome plane, while the correlation length of the $LLL$ phase is around 100 unit cells within the kagome plane and around 25 unit cells across the kagome plane. This indicates that the in-plane phase coherence of both phases is larger than their out-of-plane phase coherence, which is consistent with the out-of-equilibrium behavior where the out-of-plane stacking is more perturbed than the in-plane $2\cross2$ distortions by a light excitation. Additionally, the $MLL$ phase exhibits a longer correlation length than the $LLL$ phase.

\begin{figure*}[th!]
\includegraphics[width=0.95\textwidth]{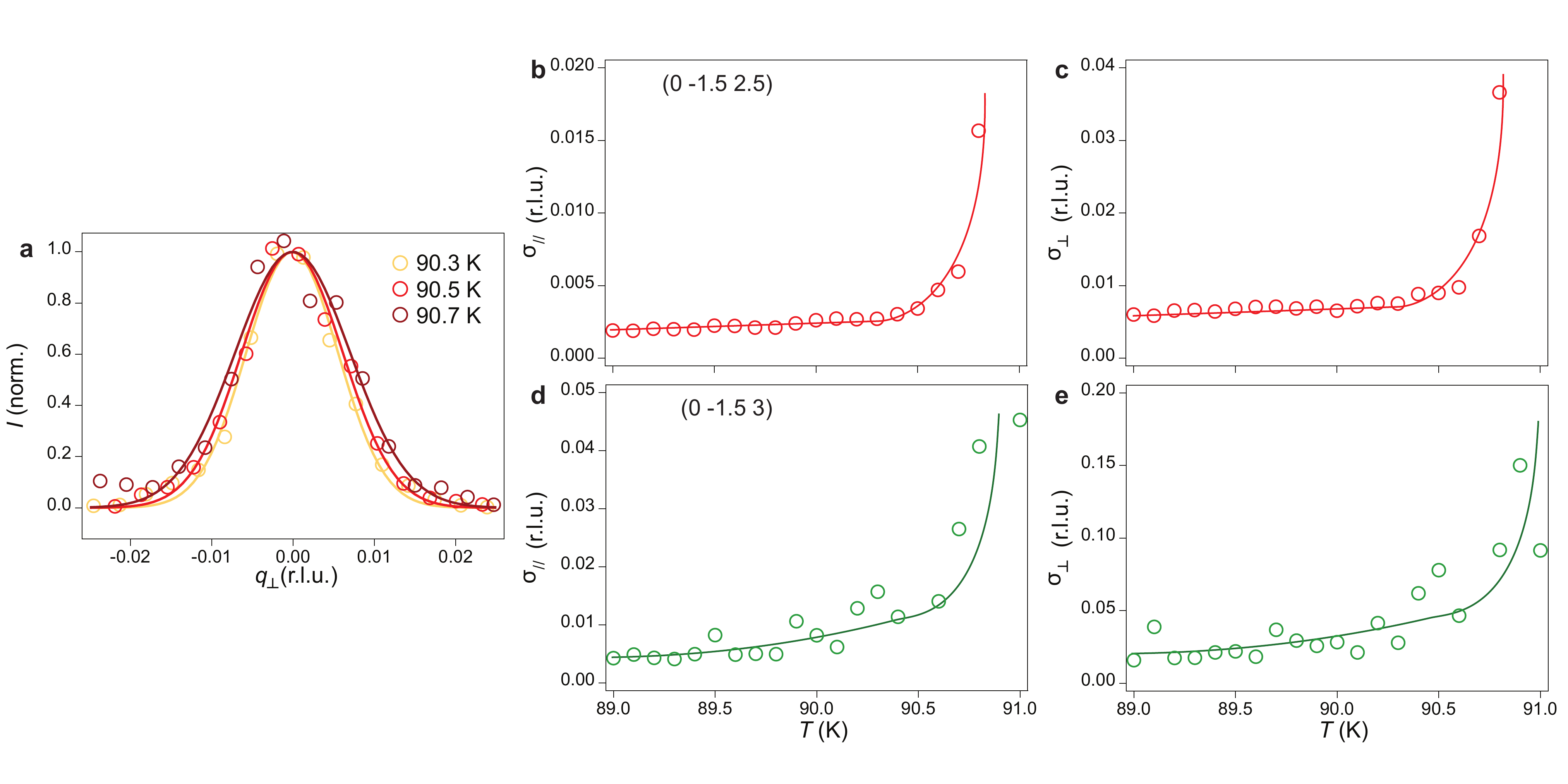}
\label{FigS11}
\caption{\textbf{Equilibrium temperature dependence of CDW peak width near $T_c$.} \textbf{a}, Integrated linecuts of the peak at (0 -1.5 2.5) along $q_{\perp}$ direction at various temperatures near $T_c$= 91 K. Solid lines are fits to a Gaussian. \textbf{b,c}, Temperature dependence of the CDW peak HWHM at (0 -1.5 2.5) along the in-plane and out-of-plane directions, respectively. \textbf{d,e}, Temperature dependence of the CDW peak HWHM at (0 -1.5 3) along the in-plane and out-of-plane directions, respectively. Solid lines are guides to eyes.} 
\end{figure*}

\section*{S10. Temporal evolution of the CDW peak position}

\begin{figure*}[th!]
\includegraphics[width=0.85\textwidth]{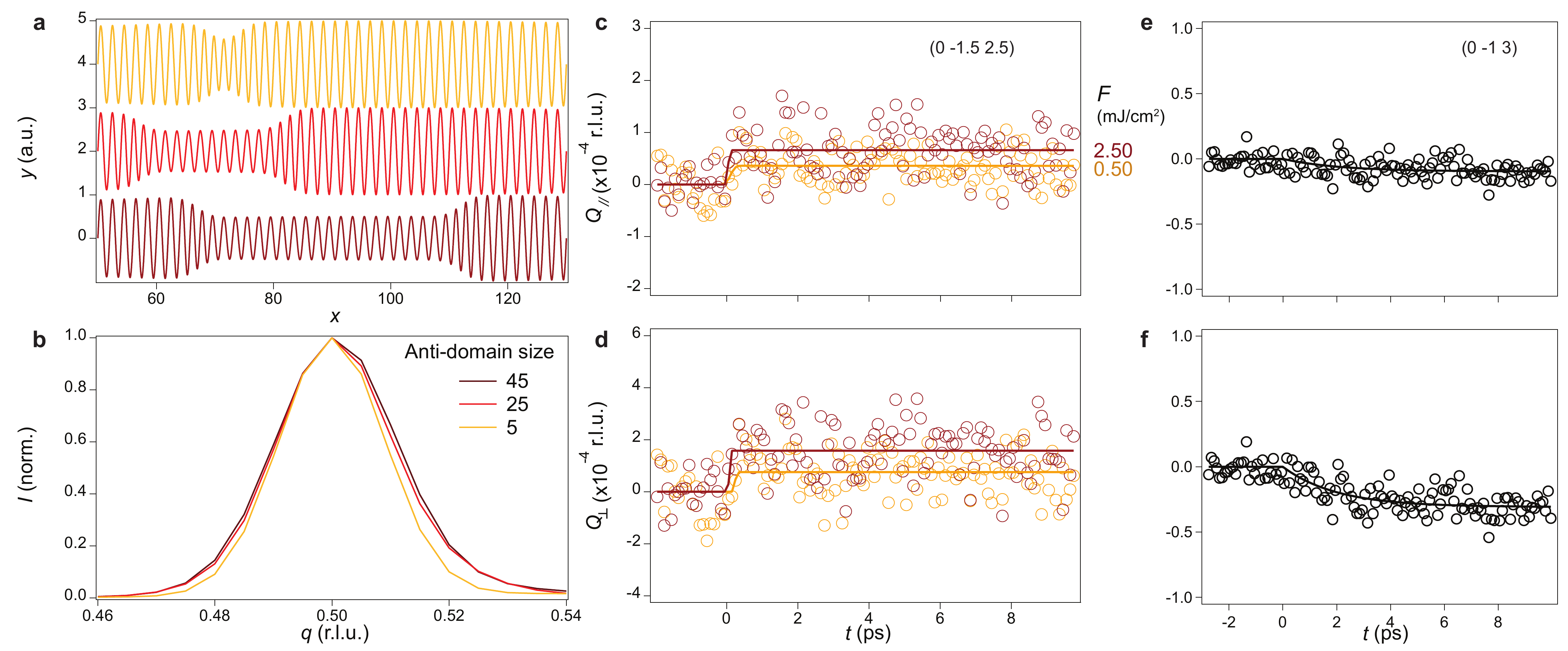}
\label{FigS12}
\caption{\textbf{Temporal evolution of the CDW peak position.} \textbf{a,} Real-space configuration of the charge density with different sizes of anti-domain. Only a subset of the simulated region is shown for clarity (50 $< x <$ 130). \textbf{b,} Normalized $|$FFT$|^2$ spectra of the CDW with different sizes of anti-domain. \textbf{c,d,} Temporal evolution of the peak position around (0 -1.5 2.5) along the in-plane and the out-of-plane directions, respectively. The data are acquired at $F=0.5$ and 2.5 mJ/cm$^2$. Solid lines are fits to a single-exponential decay. \textbf{e,f,} Temporal evolution of the peak position around (0 -1 3) along the in-plane and the out-of-plane directions, respectively. The data are acquired at $F$=2.5 mJ/cm$^2$. Solid lines are guides to eyes.} 
\end{figure*}

We observe a subtle but noticeable peak position shift accompanied by peak width modulation, manifested as an asymmetric dip-peak-dip or peak-dip-peak feature in the differential linecuts shown in Fig. 3 in the main text. Note that not only is this shift almost instantaneous upon light excitation, in contrast to the slow modulation of Bragg peak position that reflects thermal expansion, but its amplitude is also nearly one order of magnitude larger than the thermal expansion in the same timescale (Figs. S12c-f). The peak shift at first glance may be counter-intuitive because in a commensurate CDW system, the charge order is strongly pinned to the lattice, making its periodicity highly robust against perturbation. However, as we show later in a toy model simulation, the peak position change does not necessarily indicate a reciprocal lattice vector change. Instead, a change in the CDW phase can modulate the peak position in the reciprocal space.

To illustrate this effect, we conduct a toy model simulation starting with a one-dimensional CDW order with a uniform phase $\phi=0$. The order is present in a finite chain of 200 sites with a periodicity of 2 sites and a correlation length of 100 sites, mimicking the case of CsV$_3$Sb$_5$. We can express the real-space charge density modulation $y$ as a Gaussian-enveloped sinusoid, i.e. $y(x)=A\sin(2\pi Qx+\phi)\exp[-\frac{(x-\frac{d}{2})^2}{2\sigma}]$, where $Q=0.5$ is the reciprocal periodicity, $d=200$ is the size, $\sigma=100$ is the correlation length, $A=1$ is the amplitude, and $\phi=0$ is the phase. FFT indeed produces a peak centered at $q=0.5$ with a finite width proportional to the inverse of $\sigma$.

Since the periodicity is 2, the CDW phase can be either 0 or $\pi$. Now without changing the periodicity, we introduce an anti-domain, i.e., a CDW domain with the same periodicity $Q=0.5$ and opposite phase $\phi=\pi$. In the anti-domain, the phase increases from 0 to $\pi$ at both edges in 2 sites but remains at $\pi$ in the middle. We set the amplitude of the anti-domain to be half of the zero-phase domain and set the size of the anti-domain to be a value smaller than the correlation length of the dominant $\phi=0$ CDW. Therefore, effectively both $A$ and $\phi$ are functions of $x$ in the 200 sites. The presence of these domains can be visualized in real space (Fig. S12a). Then we randomly set the position of this anti-domain within the 200 sites and calculate the FFT spectra. Further, we repeat this procedure 200 times to emulate the randomness of the anti-domains distributed in real space and sum over all the FFT spectra. Finally, we convolve the total spectrum with a Gaussian function representing the momentum resolution of the experimental instrument. The $|\text{FFT}|^2$ spectra of the anti-domains of different sizes are shown in Fig. S12b. Unequivocally, as the anti-domain size increases, both the peak width increase and a shift in the peak position away from $Q=0.5$ can be observed, manifested as an asymmetric broadening. Note that this conclusion can be extended to higher-dimensional cases, and the results of the simulation will be qualitatively identical.

This toy model provides a qualitative demonstration that the modulation of the phase and amplitude of a CDW without changing its periodicity can lead to a shift in the corresponding momentum peak position and width. To be specific, the annihilation of $\phi=\pi$ domains within the dominant $\phi=0$ region will induce a peak sharpening and position shift. In the case of CsV$_3$Sb$_5$, $LLL$ surrounded by $MLL$ can be considered as the anti-domain within the $\pi=0$ domain. Therefore, along with the light-induced modulation of domain size, the peak position corresponding to different CDW phases will show a slight shift. Our simulation predicts that as the size of the anti-domain changes from 45 to 25 in the 200 sites, the peak width decreases by 3\% and the peak position shifts by 0.0003 r.l.u. (Fig. S12b). Both values qualitatively match the experimental results, demonstrating that the two domains are mesoscopically separated rather than inhomogeneously entangled in real space (Fig. S12 and Fig. 3) \cite{CoslovichScience2022}. In addition, we indeed observe a larger peak shift along $q_{\perp}$ than $q_{//}$ (Fig. S12c,d), in agreement with the larger change in peak width along $q_{\perp}$.

\section*{S11. Landau theory for two competing phases}
We consider a system with two spacetime-dependent real order parameter fields $\Delta_1(r,t)$ and $\Delta_2(r,t)$. The general free energy functional $F$ can be expressed as \cite{MillisPRX2020,MillisPRB2020}:
\begin{equation}\label{FreeEnergy2}
    F[\Delta_1,\Delta_2]\propto\int d^Dr (f_1[\Delta_1]+f_2[\Delta_2]+f_c[\Delta_1,\Delta_2]),
\end{equation}
where $f_{1,2}$ characterizes the free energy of individual $\Delta_{1,2}(r,t)$, and $f_c$ expresses the competition between the two orders where the presence of one would suppress the other such that the only stable minima are located at the $\Delta_1(r,t)$ and $\Delta_2(r,t)$ axes, i.e. one of them is zero. The free energy of both orders is assumed to be of first-order without loss of generality:
\begin{equation}
    f_i[\Delta_i]=-\alpha_i(t)\Delta_i^2+\beta_i\Delta_i^3+u_i\Delta_i^4+(\xi_{i}\nabla\Delta_i)^2.
\end{equation}
Here $\alpha_i$ and $u_i$ are positive while $\beta_i$ is negative so the minima are realized when $\Delta_i>0$. The cubic term makes the transition first-order (for the second order case, $\beta_i=0$). $\xi_i>0$ represents the correlation length. Only the quadratic term is modulated by light. The competition term has the form:
\begin{equation}
    f_c[\Delta_1,\Delta_2]=c\Delta_1^2\Delta_2^2,
\end{equation}
where $c$ is large and positive.

Based on the aforementioned assumptions, the equilibrium global minima will be reached at
\begin{equation}
    \Delta_{i,0}=\frac{-3\beta_i+\sqrt{9\beta_i^2+32u_i\alpha_i}}{8u_i}.
\end{equation}

Minimizing $f$ in one dimension with the boundary conditions $\Delta_1(x\rightarrow-\infty)=\Delta_{1,0}$, $\Delta_1(x\rightarrow\infty)=0$, $\Delta_2(x\rightarrow-\infty)=0$, and $\Delta_2(x\rightarrow\infty)=\Delta_{2,0}$, the domain wall profile can be approximated as \cite{MillisPRB2020}: 
\begin{equation}
    \Delta_{i}[x]=\Delta_{i,0}(\pm\tanh{(x/\xi_{DW})}+1)/2,
\end{equation}
where $+/-$ corresponds to $i=2$ and 1, respectively, and $\xi_{DW}\propto\xi/\sqrt{|\alpha|}$ characterizes the domain wall size.

We can also estimate the motion of interface between the two phases upon light excitation \cite{MillisPRB2020}. In the linear response regime, the displacement of a domain wall can be expressed as:
\begin{equation}
    \Delta x=\frac{1}{d_0^2}\int dt[-\Delta^2_{1,0}(\alpha_1(t)-\alpha_1(0))+\Delta^2_{2,0}(\alpha_2(t)-\alpha_2(0))],
\end{equation}
where $d_0^2\approx \frac{\alpha_1(0)}{u_1\gamma_1\xi_1}+\frac{\alpha_2(0)}{u_2\gamma_2\xi_2}$ and $\alpha(t)$ are given by Eq.S10. 

\textcolor{black}{This theory can be in principle used to capture the dynamics of our targeted system, where $\Delta_1(r,t)$ and $\Delta_2(r,t)$ represent the $MLL$ and $LLL$ phases, respectively, with the latter being more suppressed by light excitation. However, we need to treat each phase ($MLL$ and $LLL$) described by a single order parameter, which contradicts with the general Landau theory employed in this system. Furthermore, all Landau parameters as defined before remain undetermined. Given the considerable uncertainty and arbitrariness, we refrain from modeling the competition between the two phases, as we deem the experimental observation sufficiently direct evidence. }

\section*{S12. Simulated and measured fluence dependence of the CDW peak intensity}

In this section, we clarify the correspondence between the simulated and measured pump fluence dependence of the CDW diffraction peak intensity. We first show the simulation results. When the topmost single layer is 100\% melted ($F$ $\sim$ 1.7, Fig. S13a), an $\sim$18\% decrease of the integrated intensity across the probed region is reached (Fig. S13b). Since the pump has a finite penetration depth ($\sim$85 unit cells), further increasing the fluence leads to the complete melting of additional layers beneath the topmost layer. Therefore, the integrated intensity within the probe penetration depth continues to decrease (Fig. S13b). Specifically, the observed 22\% decrease signifies the complete suppression of the $LLL$ phase in the $\sim$25 top layers at our highest experimental fluence. Also note that due to the significant pump-probe penetration depth mismatch, the integrated intensity does not exhibit a distinct saturation behavior as in the topmost layer when $LLL$ is fully suppressed. Instead, only a change in photo-susceptibility (slope of $I_{int}$ vs $F$) can be observed, as captured by our experiments (Fig. 2f and Fig. S14).

\begin{figure}[th!]
\includegraphics[width=0.85\columnwidth]{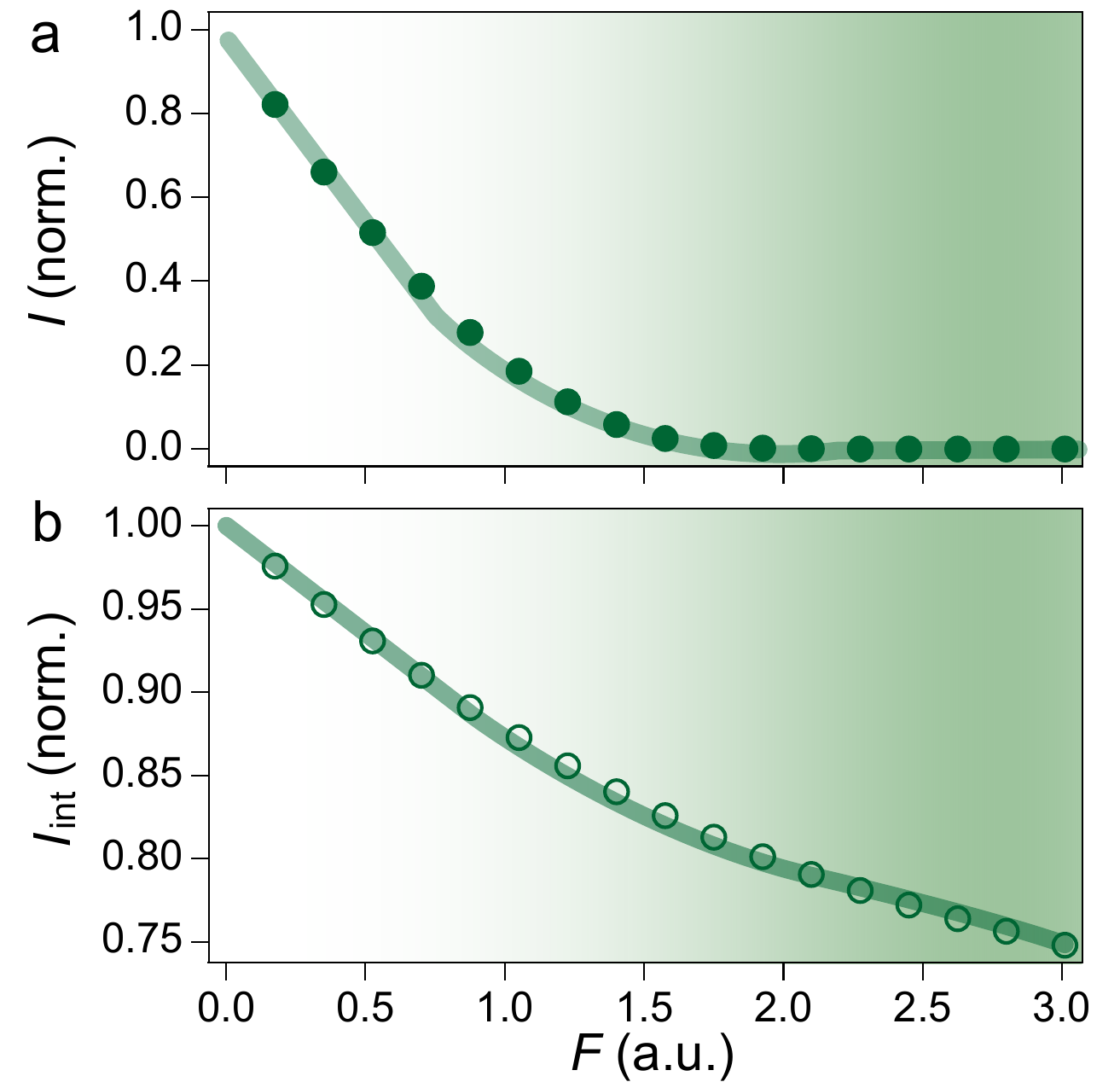}
\label{FigS13}
\caption{Pump fluence dependence of the \textbf{a}, topmost layer intensity and \textbf{b}, integrated intensity within the probe penetration depth acquired at $t$ = 0.5 ps from TDLT simulation of $LLL$. Thick lines are guides to the eyes. Shaded area denotes the region where the CDW of the topmost layer has been fully melted.}
\end{figure}

To experimentally examine the difference in photo-susceptibility between different order parameters $M$ and $L$ and different phases $MLL$ and $LLL$, we conducted a finer pump fluence dependent measurement of the three peaks at $t = 0.5$ ps, where the intensity drop is maximized. Fig. S14 clearly shows that the photo-susceptibility of the (-0.5 -1 2) peak is $\sim$3 times smaller than the (0 -1.5 2.5) peak, corroborating the $\sim3$ times higher photo-susceptibility of $L$ than $M$. Notably, both peaks do not exhibit clear saturation within the investigated fluence range. We refrain ourselves from reaching higher fluence to avoid sample damage, but an extrapolation would predict the saturation fluence to be higher than $\sim$5 mJ/cm$^2$. On the contrary, the (0 -1.5 3) peak originating from LLL shows a $\sim$2 times faster suppression and displays saturation behavior at $\sim$2 mJ/cm$^2$ when the relative change is $\sim$18\%, consistent with the results in Fig. 2 and our simulation.

\begin{figure}[th!]
\includegraphics[width=0.85\columnwidth]{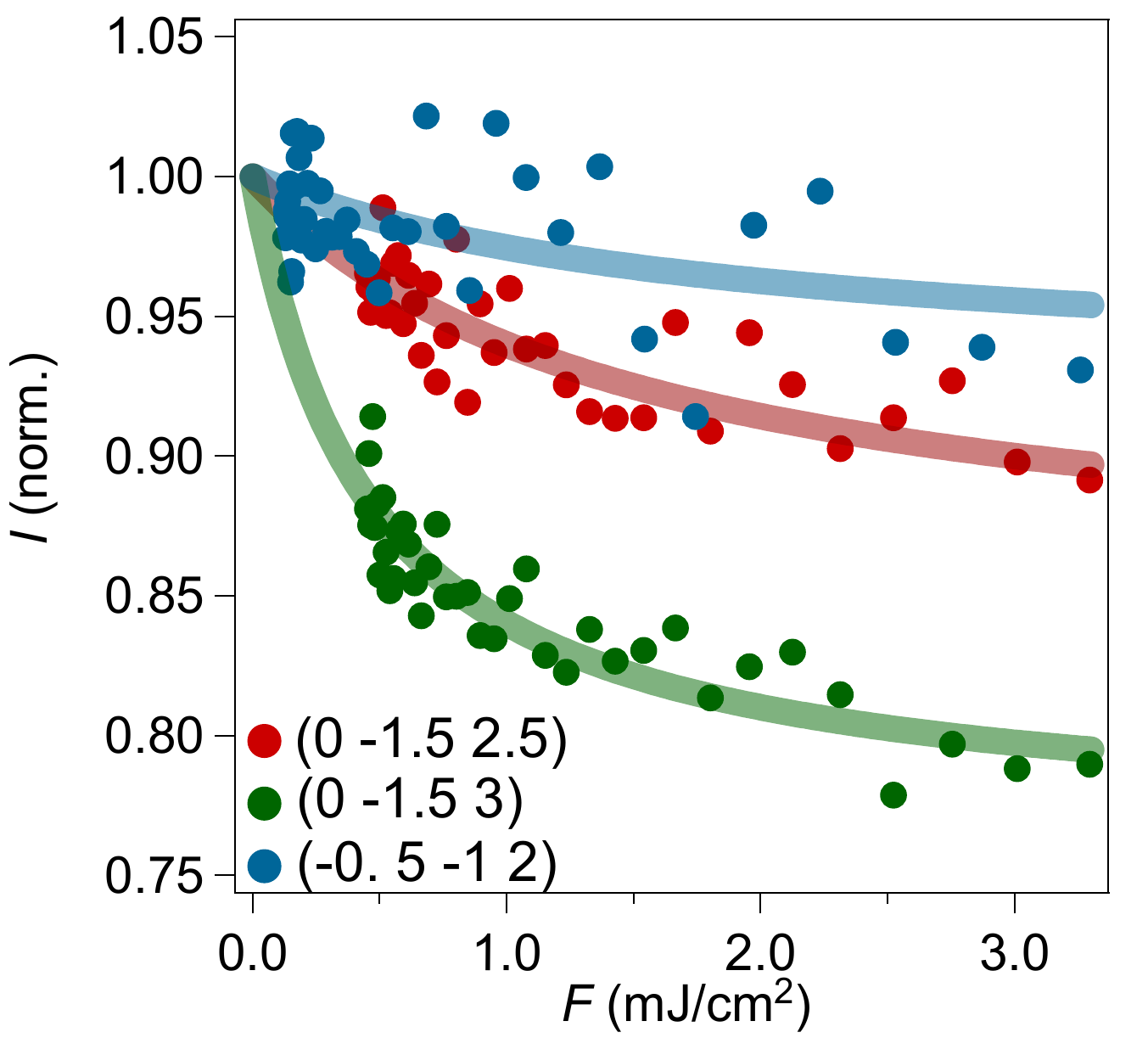}
\label{FigS14}
\caption{Fluence dependence of the intensity drop of the three CDW peaks measured at $T$ = 30 K and $t$ = 0.5 ps. Thick colored lines are guides to the eyes.} 
\end{figure}

\section*{S13. Ruling out alternative possible interpretations}

\subsection*{Mismatch in pump and probe penetration depths}

It is worth noting that the probe penetration depth varies when we measure different peaks due to our experimental geometry. We used a six-axis diffractometer with $\theta$ (or $\eta$ depending on the convention) = 2.95$^\circ$ \cite{You1999}, where the angle $\mu$, which is close to but not exactly equal to the azimuthal rotation $\phi$, is rotated to reach different diffraction peaks. Thus, as the sample is rotated, the incident angle $\beta_{in}$ is not exactly equal to $\theta$ = 2.95$^\circ$, and thus the X-ray penetration depth is not constant. Therefore, it is possible that the disparity in penetration depths when measuring different peaks gives rise to the difference in the magnitude of peak intensity drop.

To quantitatively address this effect, we calculated the real X-ray incidence angles and corresponding penetration depths for the three CDW peaks: (0, -1.5, 2.5) peak: $\beta_{in}$ = 2.94$^\circ$, $\delta_{pr}$ = 564 nm; (-0.5, -1, 2) peak: $\beta_{in}$ =  2.78$^\circ$, $\delta_{pr}$ = 534 nm; (0, -1.5, 3) peak: $\beta_{in}$ =  2.89$^\circ$, $\delta_{pr}$ = 555 nm. We find a $\sim$5\% variation in X-ray penetration depth. Precisely, the intensity drops with accurate $\delta_{pr}$ values are 0.1\%, 0.4\%, and 0.3\% larger than the intensity drops with an identical $\delta_{pr}$ = 568 nm, corresponding to a constant $\beta_{in}=\theta=2.95^\circ$, for the (0 -1.5 2.5), (-0.5 -1 2), and (0 -1.5 3) peaks, respectively. These disparities cannot account for the difference in the measured intensity drop of different peaks, which are at least one order of magnitude larger. To induce an ``artificial" difference solely from penetration depth mismatch between different peaks, a $\sim$55\% variation in x-ray penetration depth would be required, which is impractically large. Therefore, we can confidently rule out the variation in $\beta_{in}$ as the primary cause of the disparity in the amount of melting of different peaks.

The optical pump penetration depth could also vary as we rotate $\mu$, if the sample is anisotropic in plane. Here, we show that the optical in-plane anisotropy of CsV$_3$Sb$_5$ is negligible for our measurement. First, the space groups of the $MLL$ and $LLL$ phases are $Fmmm$ (point group $D_{2h}$) and $P6/mmm$ (point group $D_{6h}$), respectively. The symmetry of the latter necessitates strict in-plane isotropy of the optical constant, while the former does not. However, previous Raman spectroscopy measurements on $A_g$ and $E_{2g}$ modes indicate no polarization-angle dependence both above and below $T_c$ \cite{XXXNCOMM2022}. Additionally, no reports of in-plane anisotropy have emerged from optical conductivity measurements \cite{uykur2021low,WHHPRB2021}. These findings suggest negligible optical in-plane anisotropy, if any, when both phases coexist. Furthermore, we only moderately rotate $\mu$ by $\sim$15$^\circ$ to reach the three peaks. Therefore, the change in pump penetration depth due to the in-plane anisotropy is negligible when measuring the three peaks.

\subsection*{Static-strain induced spatial inhomogeneity of CDW domains}

Assuming the coexistence of two CDW domains ($MLL$ and $LLL$) and a static strain around the sample surface induced by cleavage, the CDW domain sizes around the surface can deviate from those in the deeper bulk. Since the surface domains are predominantly melted by light, the observed diffraction would primarily stem from the less perturbed bulk and the melting of only surface layers could result in a change in the measured peak width, which seems to explain the anomalous peak width alteration in our data. However, we provide four perspectives against this scenario as follows.

First, we note that the interlayer interaction in CsV$_3$Sb$_5$ is rather weak. We can easily exfoliate a few top layers off the sample with tape, instead of cleaving conventional 3D bulk samples by knocking off a post glued on the sample surface with epoxy. The weak interlayer coupling leads to inefficient strain transfer, and thus the strain induced by exfoliation, if any, should affect only a few layers near the sample surface. Moreover, investigations on $A$V$_3$Sb$_5$ using surface-sensitive techniques like angle-resolved photoemission spectroscopy (ARPES) and scanning tunneling microscopy (STM) show no evidence that cleavage alters the CDW distribution or structure, although cleavage has been routinely employed in these measurements.

Second, to more accurately assess the strain generated by exfoliation, we performed a toy-model calculation as illustrated in Fig. S15. In the mechanical exfoliation process, normal ($F_{\perp}$) and shear forces ($F_{//}$) are applied to the sample surface. With an order of magnitude estimate, both forces should not be over $\sim$ 1 N. These forces will thus induce a pressure $P$ = $F/A$ $\sim$ 1 N/(2 mm$\cross$2 mm) $\sim$ 0.0002 GPa, resulting in lattice deformation both perpendicular ($\Delta L$) and parallel ($\Delta x$) to the sample surface. The magnitudes of the perpendicular ($\Delta L/L$) and parallel ($\Delta x/L$) strains are determined by the Young’s modulus ($E$) and shear modulus ($G$) of the sample, which are around 89 and 33 GPa, respectively, based on DFT calculations \cite{NAHER2023106742}. These values yield $\Delta L/L$ = 0.0002\% and $\Delta x/L$ = 0.0006\%, respectively. Given $L$ $\sim$ 50 $\mathrm{\mu}$m, we have $\Delta L$ $\sim$ 1 Å and $\Delta x$ $\sim$ 3 Å. These values are even smaller than the dimension of one unit cell, which can hardly induce changes in the CDW domain sizes. Even if we assume distortion occurs entirely in one domain, e.g. $LLL$, the relative change in domain size due to exfoliation along the two directions should be $\Delta L/\sigma_{\perp}$ $\sim$ 1 Å/23 nm = 0.4\% and $\Delta x/\sigma_{//}$ $\sim$ 3 Å/55 nm = 0.5\%, respectively. If light does not modulate the domain sizes of different CDW structures but merely melt the topmost domains, the expected change in the integrated peak width $\sigma_{\perp}$ and $\sigma_{//}$ within the probed region ($\sim$560 nm) would be both around 0.01\%. This is at least two orders of magnitude smaller than our measured values ($\sim$12\% and 3\%, respectively). Consequently, the strain generated in the exfoliation process alone cannot account for the orders of magnitude larger peak width changes. 

\begin{figure*}[tbh!]
\includegraphics[width=0.85\textwidth]{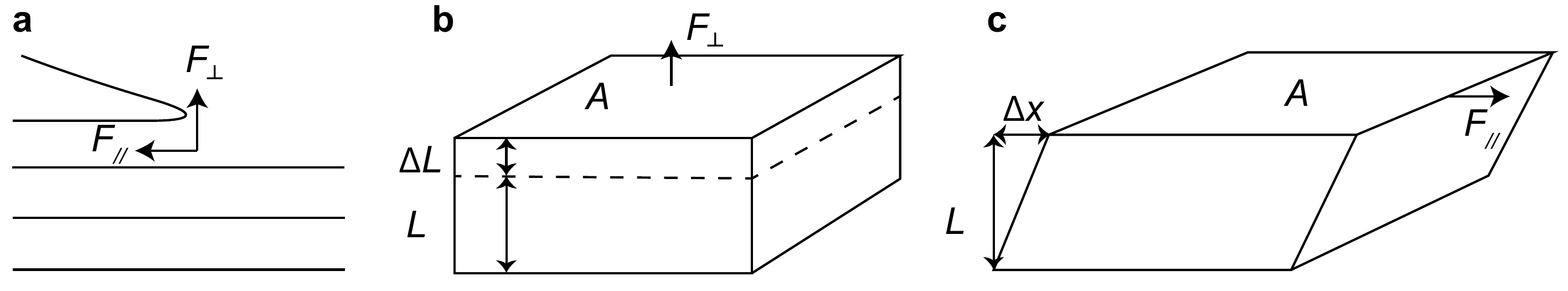}
\label{FigS15}
\caption{Illustration of \textbf{a}, the forces generated in the exfoliation process, \textbf{b}, Young’s modulus, and \textbf{c}, shear modulus.}
\end{figure*}

For reference, previous experiments on CsV$_3$Sb$_5$ that investigated the change in $T_{c}$ as a function of hydrostatic pressure \cite{yu2021unusual, chen2021double, yu2022pressure} and uniaxial strain \cite{qian2021revealing} also confirmed the negligible change in CDW induced by such a small strain/pressure. Specifically, with a pressure of 0.0002 GPa and 0.0006\% strain, the anticipated relative changes in $T_{c}$ should be around 0.001 K/91 K $\sim$ 0.001\% and 0.0001 K/91 K $\sim$ 0.0001\%, respectively.

Third, the temporal evolution of the peak width also provides evidence against this scenario. If the peak width change simply arises from the melting of the surface layers with different domain sizes, a synchronized change in peak intensity and width would be anticipated. To interrogate the initial changes in peak width and intensity, we closely examine their dynamics around time zero as shown in Fig. S16. We note that although the intensity decreases within $\sim$0.2 ps upon light excitation, potentially constrained by the time resolution, the peak width along both $q_{//}$ and $q_{\perp}$ directions shows a notable delay which continues to increase until $\sim$0.7 ps, a duration not limited by the temporal resolution. This temporal disparity unequivocally contradicts the proposed scenario.

\begin{figure}[tbh!]
\includegraphics[width=0.85\columnwidth]{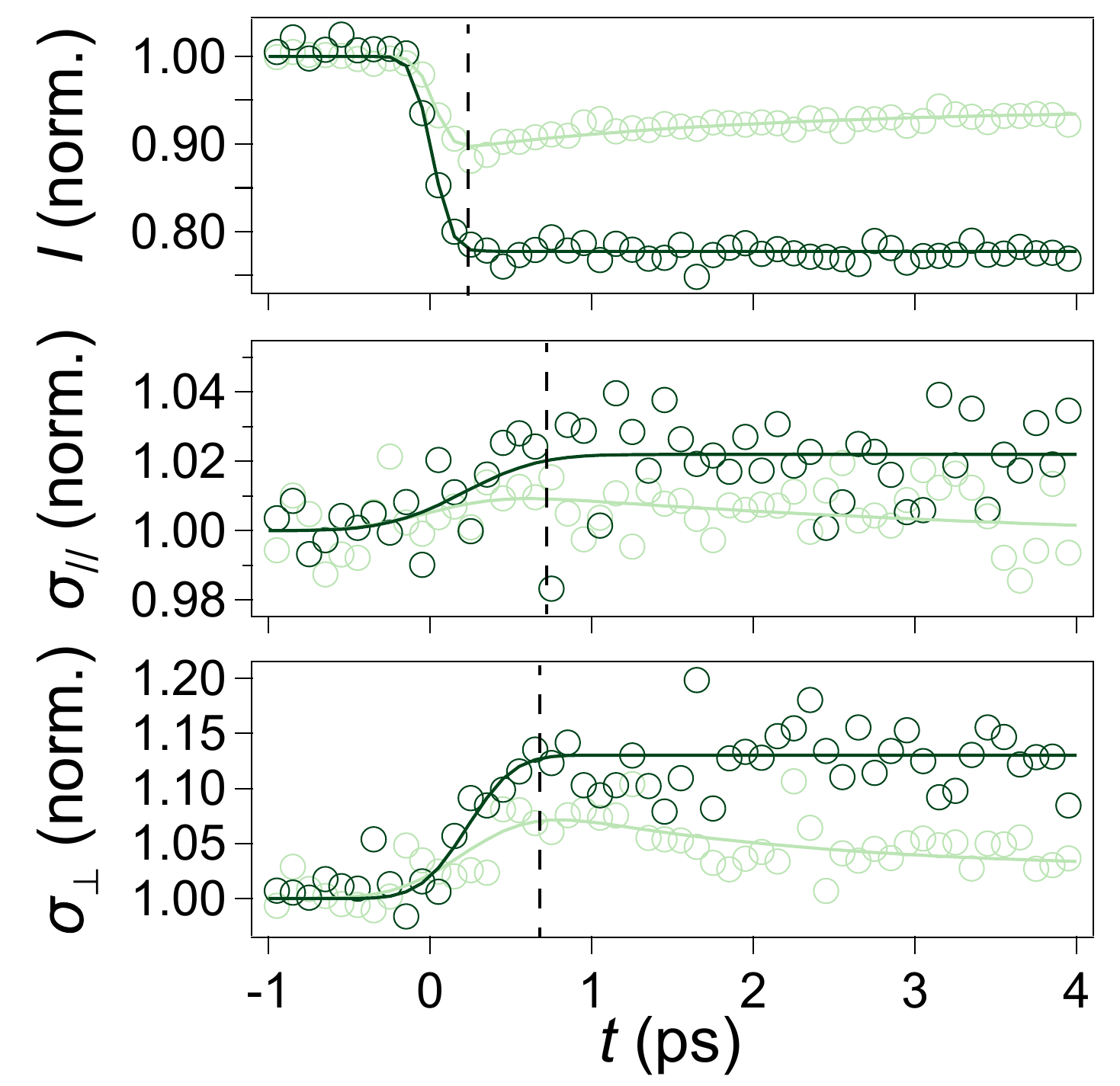}
\label{FigS16}
\caption{Temporal evolution of the intensity and the width along $q_{//}$ and $q_{\perp}$ directions of the (0 -1.5 3) peak pumped at $F$ = 0.25 (light green) and 2.5 mJ/cm$^2$ (dark green). Solid colored lines are fits to a single exponential decay. The vertical dashed lines denote the time delay when the transient changes of each variable starts to saturate or decay.} 
\end{figure}

Last, we need to note that, although the relative changes in peak width along both $q_{//}$ and $q_{\perp}$ directions for the $LLL$ peak are substantial, $\sim$3\% and $\sim$12\%, respectively, the absolute domain wall motion is not extensive, due to the relatively small static domain size of $LLL$, as demonstrated in Supplementary Section 9. The absolute domain wall motion along the two directions are $\Delta x = 100\cross5.5$ \AA$\cross$3\% = 16.5 \AA\:and $\Delta z = 25\cross9$ \AA$\cross$12\% = 27 \AA. The speed of expansion or contraction in domain size should be limited by the speed of sound. Inelastic X-ray scattering measurements and $ab$ $initio$ calculations have reported the speed of sound in CsV$_3$Sb$_5$ in  a range from 2300 to 4600 m/s \cite{MHPRX2021,NAHER2023106742}. Taking the mean value of 3400 m/s, the estimated time required for the domain wall motion along the two directions is $\sim$0.5 ps and $\sim$0.8 ps, respectively, qualitatively consistent with our experimental results as shown in Fig. S16. Therefore, our proposed mechanism of light-induced contraction of $LLL$ domain and expansion of $MLL$ domain indeed provides a persuasive explanation for our data. 

Considering these perspectives collectively, we believe that our proposed scenario of phase competition better aligns with our experimental observations. However, due to the limited signal-to-noise ratio of the peak width dynamics, we refrain from reaching quantitative conclusions about the exact timing of peak width changes in the main text.

\newpage

\providecommand{\noopsort}[1]{}\providecommand{\singleletter}[1]{#1}%

\end{document}